\documentclass[preprint]{aastex}

\usepackage{natbib}
\usepackage{rotating}
\usepackage{subfigure}

\shorttitle{VHE $\gamma$-ray Emission from 1ES\,0229+200} 
\shortauthors{E.~Aliu et al. (VERITAS collaboration)}

\begin{document}

\received{06/25/2013}
\accepted{12/21/2013}

\title{A Three-Year Multi-Wavelength Study of the Very High Energy
  $\gamma$-ray Blazar 1ES\,0229+200}

\author{
E.~Aliu\altaffilmark{1},
S.~Archambault\altaffilmark{2},
T.~Arlen\altaffilmark{3},
T.~Aune\altaffilmark{3},
B.~Behera\altaffilmark{4},
M.~Beilicke\altaffilmark{5},
W.~Benbow\altaffilmark{6},
K.~Berger\altaffilmark{7},
R.~Bird\altaffilmark{8},
A.~Bouvier\altaffilmark{9},
J.~H.~Buckley\altaffilmark{5},
V.~Bugaev\altaffilmark{5},
K.~Byrum\altaffilmark{10},
M.~Cerruti\altaffilmark{6,**},
X.~Chen\altaffilmark{11,4},
L.~Ciupik\altaffilmark{12},
M.~P.~Connolly\altaffilmark{13},
W.~Cui\altaffilmark{14},
C.~Duke\altaffilmark{15},
J.~Dumm\altaffilmark{16},
M.~Errando\altaffilmark{1},
A.~Falcone\altaffilmark{17},
S.~Federici\altaffilmark{4,11},
Q.~Feng\altaffilmark{14},
J.~P.~Finley\altaffilmark{14},
H.~Fleischhack\altaffilmark{4},
P.~Fortin\altaffilmark{6},
L.~Fortson\altaffilmark{16},
A.~Furniss\altaffilmark{9},
N.~Galante\altaffilmark{6},
G.~H.~Gillanders\altaffilmark{13},
S.~Griffin\altaffilmark{2},
S.~T.~Griffiths\altaffilmark{18},
J.~Grube\altaffilmark{12},
G.~Gyuk\altaffilmark{12},
D.~Hanna\altaffilmark{2},
J.~Holder\altaffilmark{7},
G.~Hughes\altaffilmark{4},
T.~B.~Humensky\altaffilmark{19},
C.~A.~Johnson\altaffilmark{9},
P.~Kaaret\altaffilmark{18},
M.~Kertzman\altaffilmark{20},
Y.~Khassen\altaffilmark{8},
D.~Kieda\altaffilmark{21},
H.~Krawczynski\altaffilmark{5},
F.~Krennrich\altaffilmark{22},
M.~J.~Lang\altaffilmark{13},
A.~S~Madhavan\altaffilmark{22},
G.~Maier\altaffilmark{4},
P.~Majumdar\altaffilmark{3,23},
S.~McArthur\altaffilmark{24},
A.~McCann\altaffilmark{25},
K.~Meagher\altaffilmark{26},
J.~Millis\altaffilmark{27},
P.~Moriarty\altaffilmark{28},
R.~Mukherjee\altaffilmark{1},
D.~Nieto\altaffilmark{19},
A.~O'Faol\'{a}in de Bhr\'{o}ithe\altaffilmark{8},
R.~A.~Ong\altaffilmark{3},
A.~N.~Otte\altaffilmark{26},
N.~Park\altaffilmark{24},
J.~S.~Perkins\altaffilmark{29,*},
M.~Pohl\altaffilmark{11,4},
A.~Popkow\altaffilmark{3},
H.~Prokoph\altaffilmark{4},
J.~Quinn\altaffilmark{8},
K.~Ragan\altaffilmark{2},
L.~C.~Reyes\altaffilmark{30},
P.~T.~Reynolds\altaffilmark{31},
G.~T.~Richards\altaffilmark{26},
E.~Roache\altaffilmark{6},
G.~H.~Sembroski\altaffilmark{14},
A.~W.~Smith\altaffilmark{21},
D.~Staszak\altaffilmark{2},
I.~Telezhinsky\altaffilmark{11,4},
M.~Theiling\altaffilmark{14},
A.~Varlotta\altaffilmark{14},
V.~V.~Vassiliev\altaffilmark{3},
S.~Vincent\altaffilmark{4},
S.~P.~Wakely\altaffilmark{24},
T.~C.~Weekes\altaffilmark{6},
A.~Weinstein\altaffilmark{22},
R.~Welsing\altaffilmark{4},
D.~A.~Williams\altaffilmark{9},
A.~Zajczyk\altaffilmark{5},
B.~Zitzer\altaffilmark{10}
}

\altaffiltext{1}{Department of Physics and Astronomy, Barnard College, Columbia University, NY 10027, USA}
\altaffiltext{2}{Physics Department, McGill University, Montreal, QC H3A 2T8, Canada}
\altaffiltext{3}{Department of Physics and Astronomy, University of California, Los Angeles, CA 90095, USA}
\altaffiltext{4}{DESY, Platanenallee 6, 15738 Zeuthen, Germany}
\altaffiltext{5}{Department of Physics, Washington University, St. Louis, MO 63130, USA}
\altaffiltext{6}{Fred Lawrence Whipple Observatory, Harvard-Smithsonian Center for Astrophysics, Amado, AZ 85645, USA}
\altaffiltext{7}{Department of Physics and Astronomy and the Bartol Research Institute, University of Delaware, Newark, DE 19716, USA}
\altaffiltext{8}{School of Physics, University College Dublin, Belfield, Dublin 4, Ireland}
\altaffiltext{9}{Santa Cruz Institute for Particle Physics and Department of Physics, University of California, Santa Cruz, CA 95064, USA}
\altaffiltext{10}{Argonne National Laboratory, 9700 S. Cass Avenue, Argonne, IL 60439, USA}
\altaffiltext{11}{Institute of Physics and Astronomy, University of Potsdam, 14476 Potsdam-Golm, Germany}
\altaffiltext{12}{Astronomy Department, Adler Planetarium and Astronomy Museum, Chicago, IL 60605, USA}
\altaffiltext{13}{School of Physics, National University of Ireland Galway, University Road, Galway, Ireland}
\altaffiltext{14}{Department of Physics, Purdue University, West Lafayette, IN 47907, USA }
\altaffiltext{15}{Department of Physics, Grinnell College, Grinnell, IA 50112-1690, USA}
\altaffiltext{16}{School of Physics and Astronomy, University of Minnesota, Minneapolis, MN 55455, USA}
\altaffiltext{17}{Department of Astronomy and Astrophysics, 525 Davey Lab, Pennsylvania State University, University Park, PA 16802, USA}
\altaffiltext{18}{Department of Physics and Astronomy, University of Iowa, Van Allen Hall, Iowa City, IA 52242, USA}
\altaffiltext{19}{Physics Department, Columbia University, New York, NY 10027, USA}
\altaffiltext{20}{Department of Physics and Astronomy, DePauw University, Greencastle, IN 46135-0037, USA}
\altaffiltext{21}{Department of Physics and Astronomy, University of Utah, Salt Lake City, UT 84112, USA}
\altaffiltext{22}{Department of Physics and Astronomy, Iowa State University, Ames, IA 50011, USA}
\altaffiltext{23}{Saha Institute of Nuclear Physics, Kolkata 700064, India}
\altaffiltext{24}{Enrico Fermi Institute, University of Chicago, Chicago, IL 60637, USA}
\altaffiltext{25}{Kavli Institute for Cosmological Physics, University of Chicago, Chicago, IL 60637, USA}
\altaffiltext{26}{School of Physics and Center for Relativistic Astrophysics, Georgia Institute of Technology, 837 State Street NW, Atlanta, GA 30332-0430}
\altaffiltext{27}{Department of Physics, Anderson University, 1100 East 5th Street, Anderson, IN 46012}
\altaffiltext{28}{Department of Life and Physical Sciences, Galway-Mayo Institute of Technology, Dublin Road, Galway, Ireland}
\altaffiltext{29}{N.A.S.A./Goddard Space-Flight Center, Code 661, Greenbelt, MD 20771, USA}
\altaffiltext{30}{Physics Department, California Polytechnic State University, San Luis Obispo, CA 94307, USA}
\altaffiltext{31}{Department of Applied Physics and Instrumentation, Cork Institute of Technology, Bishopstown, Cork, Ireland}

\altaffiltext{*}{Corresponding author: jeremy.s.perkins@nasa.gov}
\altaffiltext{**}{Corresponding author: mcerruti@cfa.harvard.edu}

\begin{abstract}
The high-frequency-peaked BL Lacertae object 1ES\,0229+200 is a
relatively distant ({\it z} = 0.1396), hard-spectrum ($\Gamma \sim
2.5$), very-high-energy-emitting ($E >$ 100 GeV) $\gamma$-ray
blazar. Very-high-energy measurements of this active galactic nucleus
have been used to place constraints on the intensity of the
extragalactic background light and the intergalactic magnetic field.
A multi-wavelength study of this object centered around
very-high-energy observations by VERITAS is presented.  This study
obtained, over a period of three years, an 11.7 standard deviation
detection and an average integral flux $F (E>300~{\rm GeV}) = (23.3
\pm 2.8_{\rm stat} \pm 5.8_{\rm sys}) \times 10^{-9}$ photons m$^{-2}$
s$^{-1}$, or 1.7\% of the Crab Nebula's flux (assuming the Crab Nebula
spectrum measured by H.E.S.S). Supporting observations from {\it
  Swift} and RXTE are analyzed.  The {\it Swift} observations are
combined with previously published {\it Fermi} observations and the
very-high-energy measurements to produce an overall spectral energy
distribution which is then modeled assuming one-zone
synchrotron-self-Compton emission. The $\chi^2$ probability of the TeV
flux being constant is 1.6\%.  This, when considered in
combination with measured variability in the X-ray band, and the
demonstrated variability of many TeV blazars, suggests that the use of
blazars such as 1ES\,0229+200 for intergalactic magnetic field studies
may not be straightforward and challenges models that attribute hard
TeV spectra to secondary $\gamma$-ray production along the line of
sight.
\end{abstract}

\keywords{extragalactic --- BL Lacertae objects: individual
  (1ES\,0229+200, VER\,J0232+202) --- gamma-rays: observations}

\section{Introduction}

The detection of the hard-spectrum, distant blazar
\objectname{1ES\,0229+200} at very high energies (VHE; $E > $100 GeV)
by H.E.S.S in 2007 \citep{Aharonian:2007dn} generated excitement among
the members of the VHE community, especially those members that study
the extragalactic background light (EBL) and the intergalactic
magnetic field (IGMF). It was well known that VHE $\gamma$-rays are
attenuated via pair production on the mid-infrared EBL as they
propagate through the Universe \citep{Gould:1967uf,Coppi:1998fk} and
that the pairs are then deflected by the IGMF \citep[for
  example,][]{Neronov:2009fk}. However, the majority of the models of
the EBL at the time postulated a strong EBL and a relatively nearby
$\gamma$-ray horizon \citep[for
  example,][]{Stecker:2006kx,Kneiske:2002ib}.  The discovery of a
hard-spectrum (spectral index smaller than 3.0) blazar at TeV energies
with a well-determined redshift above 0.1 cast doubt upon the strong
EBL scenario.

Distant, hard-spectrum blazars are also ideal for studies of the IGMF
for similar reasons.  The pairs produced in EBL interactions are
deflected by the IGMF before interacting with cosmic microwave
background (CMB) photons via inverse-Compton scattering (the photons
produced in the inverse-Compton scatterings off of the CMB will have
GeV energies). If the IGMF is not overly strong, the resulting
high-energy (GeV) and VHE photons are directed along the path of the
original emitted photon \citep[for a discussion of this,
  see][]{Dermer:2011fk, Taylor:2011fk, Dolag:2011uq, Vovk:2012vn,
  Arlen:2012kx}.  This effect can cause a delay in the arrival of the
signal and extended emission around point sources \citep[for a review
  of these processes see, for example,][]{Neronov:2009fk}. Significant
effort has been made to place limits on the IGMF using VHE and GeV
blazars by comparing the flux seen in the two energy bands \citep[for
  example,][]{Dermer:2011fk}.  Since the reprocessing occurs over time
(the exact time depends on the IGMF strength, coherence length and
distance to the source), these arguments usually depend on the VHE
flux not varying, at least during the period of observation.

Ever since the discovery of 1ES\,1101-232, H\,2356-309
\citep{Aharonian:2006oc} and, H\,1426+428 \citep{Horan:2002fk,
  Aharonian:2003bs} at VHE, the community has been systematically
searching for distant, hard-spectrum blazars.  The discovery of
1ES\,0229+200 was part of a series of VHE detections of active
galactic nuclei (AGN) of this class, including 1ES\,1218+304
\citep{Albert:2006uv}, 1ES\, 0347-121 \citep{Aharonian:2007vn}, and
1ES\,1101-232 \citep{Aharonian:2006oc}.  1ES\,0229+200 was especially
interesting because its measured spectrum extended up to 10
TeV\citep{Aharonian:2007dn}.  This opened up the possibility of using
such observations to study the history of the Universe, instead of
just AGN emission mechanisms \citep{Aharonian:2006oc}.  A complication
of these types of studies is that they require several conditions: a
distant source (which maximizes the attenuation length), a hard
spectral index (which increases the statistics at the highest
energies), knowledge about the intrinsic spectral index of the source,
and, specifically in the case of IGMF studies, a constant flux, so
that one can estimate the total fluence of the object over time.

The sensitivity of the current generation of VHE observatories has
allowed the detection of objects at greater redshifts.  This is
especially pertinent to the study of the EBL, which is produced from
direct and reprocessed (by dust) starlight and AGN emission
\citep{Gould:1967uf,Stecker:1992fk}. Thus, the precise measurement of
the EBL informs us of the structure formation in the Universe in early
times.  The EBL in optical to infrared wavelengths attenuates
high-energy photons through pair production \citep[$\gamma_{\rm vhe} +
  \gamma_{\rm ebl} \rightarrow e^+e^-$,][]{Gould:1967uf}.  This
directly affects the measurement of distant VHE sources by attenuating
the emitted flux and softening the spectrum.  It also effectively
places a limit on the distance accessible by $\gamma$-ray studies (the
$\gamma$-ray horizon).  The converse of this is that observations of
distant objects at high energy can be used to constrain the density of
the EBL along the line of sight to the object \citep[for example,
  see][]{Abramowski:2013qf,Ackermann:2012ly}.  This is especially
relevant as it is difficult to directly measure the EBL at the
wavelengths that affect $\gamma$-ray photons.

1ES\,0229+200 is at a redshift of {\it z} = $0.1396 \pm
0.0001$\citep{Woo:2005fk} and has an archival spectral index at VHE of
$2.50 \pm 0.19$ \citep{Aharonian:2007dn}.  These features make it
ideally suited to study the EBL. In addition, the lack of historical
evidence of VHE variability was used by some authors to justify using
measurements of 1ES\,0229+200 to constrain the strength of the IGMF
\citep{Arlen:2012kx, Dermer:2011fk, Huan:2011vn, Neronov:2010uq,
  Georganopoulos:2010ys}. The original H.E.S.S. measurement, one of
the first of its type, indicated that either the intrinsic spectral
index of the blazar was much harder than 1.5 or the EBL density in the
mid-infrared range was close to the lower limits given by {\it
  Spitzer} \citep{Fazio:2004ka} and Infrared Space Observatory data
\citep{Elbaz:2002zr} based on galaxy counts.  This measurement (along
with other contemporary measurements of blazars like 1ES\,1011-232 and
H\,2356-309) strongly disfavored many of the contemporary models of
the EBL and indicated that the $\gamma$-ray horizon was much farther
than previously thought.  Over the past several years, many other
population studies of VHE blazars have been done which corroborate
that the EBL is close to or at the lower limits
\citep{Ackermann:2012ly, Abramowski:2013qf, Orr:2011uq, Raue:2008fk}.
Similar efforts have been made to place limits on the
IGMF. \citet{Dermer:2011fk} compared the measurements by the {\it
  Fermi}-LAT with those taken at VHE and conclude that the IGMF is
very small.  Several of the EBL and IGMF studies include 1ES\,0229+200
and many of these studies depend on a long-term ($\sim$ years)
steady-state flux from the source, at least during the time that the
source is being monitored.  However, historical data have shown that
most, if not all, blazars are variable at VHE \citep[for
  example,][]{Boettcher:2010fk}.

\cite{Schachter:1993uq} identified 1ES\,0229+200 as a BL Lacertae
object after it was discovered in the Einstein IPC Slew Survey
\citep{Elvis:1992kx}.  Like most VHE blazars, it is classified as a
high-synchrotron-peaked blazar (HSP) due to the location of its
synchrotron peak \citep[as defined by][]{Ackermann:2011fk}.  As
mentioned before, it has a well measured redshift of {\it z} = 0.1396
\citep{Woo:2005fk} and is hosted by a faint elliptical galaxy
\citep[$M_{\rm R}$ = -24.53,][]{Falomo:2000rs}. As early as
\citeyear{Stecker:1996bh}, \citeauthor{Stecker:1996bh} predicted that
this HSP would emit VHE $\gamma$-rays based on its high synchrotron
peak, and \citeauthor{Costamante:2002dq} included it in their
\citeyear{Costamante:2002dq} list of possible TeV sources.  However,
the first generation of VHE instruments did not detect it
\citep{de-la-Calle-Perez:2003cr,Aharonian:2004pi, Williams:2005nx}.
When first detected by H.E.S.S in 2007 it was one of the most distant
VHE objects known at the time with spectral information at 10 TeV.
This, plus the lack of multi-wavelength observations, prompted further
study.

In this paper, we present a long-term VHE study over three seasons of
this unique blazar using the Very Energetic Radiation Imaging
Telescope Array System (VERITAS).  We investigate the repercussions of
the measurement on the EBL and IGMF and comment on this AGN's place in
the VHE blazar population.

\section{VERITAS Observations and Results}

\subsection{Observations}

VERITAS is a ground-based imaging atmospheric Cherenkov telescope
(IACT) array located at the Fred Lawrence Whipple Observatory in
southern Arizona.  The array consists of four 12 m diameter
telescopes, each with a total mirror area of 110 m$^2$.  Cameras,
located 12 m in front of the dishes, contain 499 circular
photomultiplier tubes (PMTs), yielding a field of view (FoV) of 3.5
degrees.  Winston cones are installed in front of the PMTs to reduce
the albedo and increase the light-collecting area of the camera by
filling in the gaps between them.  As an IACT, VERITAS detects the
brief flashes of Cherenkov light produced by the particle shower
induced when a $\gamma$-ray produces an electron-positron pair in the
upper atmosphere. The reconstruction of the particle shower from the
imaging of the Cherenkov flash gives the energy, time of detection and
arrival direction of the initial photon.  Overall, VERITAS can detect
photons from 100 GeV up to 30 TeV at an energy resolution of 15\% and
angular resolution smaller than 0.1 degree at 1 TeV
\citep{Holder:2008sz}.

VERITAS has a three-level trigger to reduce the rate of background
events from the night sky and local muons.  Each shower that triggers
the system is imaged by the array and stored to disk.  These shower
events are calibrated and cleaned using quality selections based on
the number of triggered photomultiplier tubes in each image and the
position of the image in the camera.  Then, the shape and orientation
of the Cherenkov images are parameterized by their principal moments
\citep{Hillas:1985vc}.  These parameters are compared to Monte Carlo
simulations of $\gamma$-ray-initiated air showers.  Cuts based on the
physical differences between $\gamma$-ray and hadronic showers, and
optimized on data taken on the Crab Nebula, are used to reject a
majority of the cosmic ray events (which are vastly more numerous than
the $\gamma$-ray showers). A bright source with a flux on the order of
10\% of the Crab Nebula's flux can be detected at a significance of 5
standard deviations ($\sigma$) in 30 minutes, while a weaker source
(1\% of the Crab Nebula's flux) can be detected in $\sim25$ hours.
More details on the VERITAS array, specifically the detection and
analysis techniques, can be found in \citet{Acciari:2008oq}.

\begin{sidewaystable}
\begin{center}
\small
  \caption{\label{tab:vtsobs}The VERITAS 1ES\,0229+200 observation
    details. $\alpha$ (the ratio of the area $\times$ livetime of the
    {\it on} source and {\it off} source regions) is 1/11.  The
    integral flux is calculated assuming an overall spectral index of
    2.59.  Upper limits at the 99\% confidence level using the Rolke
    method \citep{Rolke:2001fk} are presented when the significance is
    less than two standard deviations.  The horizontal lines delineate
    the results for the full time period, the data divided by season,
    and the data divided by observing period (dictated by the lunar
    cycle and indicated by 'P. 1' through 'P. 5' in each season).}

\begin{tabular}{cccccccccc}
  \tableline\tableline
  Period & Dates & Live Time & {\it On} & {\it Off} & Significance & Flux ( $>$ 300  GeV) & UL ($>$ 300  GeV)\\
   & [MJD] & [minutes] & [events] & [events] & [$\sigma$] & [10$^{-9}$ m$^{-2}$ s$^{-1}$] & [10$^{-9}$ m$^{-2}$ s$^{-1}$]\\
  \tableline
  2009-2012 & 55118 - 55951 & 3260 & 1917 & 15704 & 11.7 & $23.3 \pm 2.8_{\rm stat} \pm 5.8_{\rm sys}$ & N/A \\
  \tableline
  2009-2010 & 55118 - 55212 & 1674 & 1054 & 7601 & 12.2 & 30.3 $\pm 3.9_{\rm stat} \pm 7.6_{\rm sys}$ & N/A\\
  2010-2011 & 55476 - 55587 & 1079 & 614 & 5862 & 3.3 & 18.7 $\pm 5.1_{\rm stat} \pm 5.7_{\rm sys}$ & N/A\\
  2011-2012 & 55828 - 55951 & 507 & 249 & 2241 & 2.9 & 9.9 $\pm 6.4_{\rm stat} \pm 2.5_{\rm sys}$ & N/A\\
  \tableline
  2009-2010 P. 1 & 55118 - 55131 & 715 & 484 & 3210 & 9.7 & 41.8 $\pm 6.4_{\rm stat} \pm 10.5_{\rm sys}$ & N/A\\
  2009-2010 P. 2 & 55144 - 55159 & 844 & 524 & 3880 & 8.1 & 24.2 $\pm 5.4_{\rm stat} \pm 6.1_{\rm sys}$ & N/A\\
  2009-2010 P. 3 & 55183 - 55183 & 24 & 10 & 120 & -0.3 & 1 $\pm 26_{\rm stat} \pm 1_{\rm sys}$ & 100 \\
  2009-2010 P. 4 & 55200 - 55212 & 91 & 36 & 391 & 0.1 & 3 $\pm 10_{\rm stat} \pm 1_{\rm sys}$ & 51 \\
  2010-2011 P. 1 & 55476 - 55482 & 319 & 187 & 1900 & 1.0 & 15 $\pm 9_{\rm stat} \pm 4_{\rm sys}$ & 41 \\
  2010-2011 P. 2 & 55501 - 55513 & 162 & 121 & 901 & 3.8 & 39 $\pm 14_{\rm stat} \pm 10_{\rm sys}$ & N/A\\
  2010-2011 P. 3 & 55526 - 55538 & 127 & 69 & 692 & 0.7 & 1 $\pm 14_{\rm stat} \pm 1_{\rm sys}$ & 60 \\
  2010-2011 P. 4 & 55555 - 55570 & 297 & 147 & 1490 & 1.0 & 15 $\pm 10_{\rm stat} \pm 4_{\rm sys}$ & 40 \\
  2010-2011 P. 5 & 55583 - 55587 & 174 & 90 & 879 & 1.1 & 26 $\pm 13_{\rm stat} \pm 7_{\rm sys}$ & 54 \\
  2011-2012 P. 1 & 55828 - 55840 & 101 & 46 & 434 & 1.0 & 13 $\pm 14_{\rm stat} \pm 3_{\rm sys}$ & 66 \\
  2011-2012 P. 2 & 55855 - 55861 & 111 & 55 & 460 & 1.9 & 15 $\pm 14_{\rm stat} \pm 4_{\rm sys}$ & 78 \\
  2011-2012 P. 3 & 55886 - 55895 & 119 & 68 & 608 & 1.6 & 13 $\pm 14_{\rm stat} \pm 3_{\rm sys}$ & 77 \\
  2011-2012 P. 4 & 55916 - 55922 & 103 & 41 & 435 & 0.2 & -6 $\pm 13_{\rm stat} \pm 2_{\rm sys}$ & 51 \\
  2011-2012 P. 5 & 55940 - 55951 & 73 & 39 & 304 & 1.9 & 16 $\pm 18_{\rm stat} \pm 4_{\rm sys}$ & 100 \\
  \tableline
\end{tabular}
\end{center}
\end{sidewaystable}

The VERITAS collaboration has initiated a long-term science plan which
includes the observation of relatively distant blazars with hard
spectra.  The goal of this strategy is to build up a database of
spectral energy distributions (SEDs) from a variety of blazars whose
emission can carry the signature of the EBL it traverses and to study
the blazar population in greater detail. As part of this program,
VERITAS observed 1ES\,0229+200 for a total time of 54.3 hours from
2010 to 2012.  These observations were taken over three seasons (27.9,
18.0 and 8.5 hours in 2009-2010, 2010-2011 and 2011-2012,
respectively, after data quality selections for weather and other
issues) and resulted in a strong detection of 11.7 $\sigma$.  For
details on the observations see Table \ref{tab:vtsobs}.  VERITAS
observed this source in a wobble configuration, where the telescopes
are pointed 0.5 degrees away from the source so that a simultaneous
background sample can be taken along with the {\it on}-source
observations \citep{Fomin:1994kl}.

\subsection{Results}

These observations resulted in 1917 {\it on}-source events and 15704
{\it off}-source events.  The {\it off}-source region is larger than
the {\it on}-source region by a factor of eleven so the resulting
excess is 489 events, corresponding to a $\gamma$-ray rate of $(0.150
\pm 0.014)$ photons per minute.  This corresponds to an average
integral flux above 300 GeV of ($23.3 \pm 2.8_{\rm stat}\pm
5.8_{sys}$) $\times 10^{-9}$ photons m$^{-2}$ s$^{-1}$ or about 1.7\%
of the Crab Nebula's flux \citep[as measured by][]{Aharonian:2006id}.
On average, this is a similar flux to that seen by the
H.E.S.S. collaboration in 2005 - 2006 (1.8\% of the Crab Nebula's
flux) in 41.8 hours of observation \citep{Aharonian:2007dn}.  A
two-dimensional Gaussian fit to the VERITAS excess
(\objectname{VER\,J0232+202}) is consistent with a point source
located at R.A. = $02^{\rm h} 32^{\rm m} 48^{\rm s} \pm 2^{\rm s}_{\rm
  stat} \pm 6^{\rm s}_{\rm sys}$, DEC.  = $+20^{\circ}17'22'' \pm
23''_{\rm stat} \pm 1'30''_{\rm sys}$. This is $9.8''$ away from the
VLA position of the blazar \citep[R.A. = $02^{\rm h} 32^{\rm m}
  48^{\rm s}.6$, DEC. = $+20^{\circ}17'17''$,][]{Schachter:1993uq} and
within the VERITAS PSF.

\begin{table}
\begin{center}
  \caption{The VERITAS spectral bins. There is an additional 20\%
    systematic error on the Flux.\label{tab:vtsspectrum}}
\begin{tabular}{ccccccc}
  \tableline\tableline
  E & E$_{\rm low}$ & E$_{\rm high}$ & Flux & Flux Error & Excess & Significance \\
  $[{\rm TeV}]$ & [TeV] & [TeV] & [m$^{-2}$~TeV$^{-1}$~s$^{-1}$]& [m$^{-2}$~TeV$^{-1}$~s$^{-1}$] & & [$\sigma$] \\
  \tableline
  0.291 & 0.240 & 0.353 & $1.2\times10^{-7}$ & $2.8\times10^{-8}$ & 93 & 4.5\\
  0.427 & 0.353 & 0.518 & $6.1\times10^{-8}$ & $1.0\times10^{-8}$ & 106 & 6.7\\
  0.628 & 0.518 & 0.761 & $2.2\times10^{-8}$ & $4.4\times10^{-9}$ & 71 & 5.6\\
  0.922 & 0.761 & 1.12 & $4.3\times10^{-9}$ & $1.7\times10^{-9}$ & 25 & 2.6\\
  1.36 & 1.12 & 1.64 & $2.2\times10^{-9}$ & $8.1\times10^{-10}$ & 22 & 3.0\\
  1.99 & 1.64 & 2.41 & $1.0\times10^{-9}$ & $4.2\times10^{-10}$ & 16 & 2.7\\
  2.92 & 2.41 & 3.54 & $3.5\times10^{-10}$ & $2.1\times10^{-10}$ & 8 & 1.8\\
  4.29 & 3.54 & 5.20 & $2.4\times10^{-10}$ & $1.2\times10^{-10}$ & 9 & 2.6\\
  7.64 & 5.20 & 11.2 & $2.8\times10^{-11}$ & $2.5\times10^{-11}$ & 4 & 1.2\\
  16.3 & 11.2 & 24.2 & $1.4\times10^{-11}$ & {\it 99\% upper limit} & -0.7 & -0.6\\ 
  \tableline
\end{tabular}
\end{center}
\end{table}

\begin{figure}
\plotone{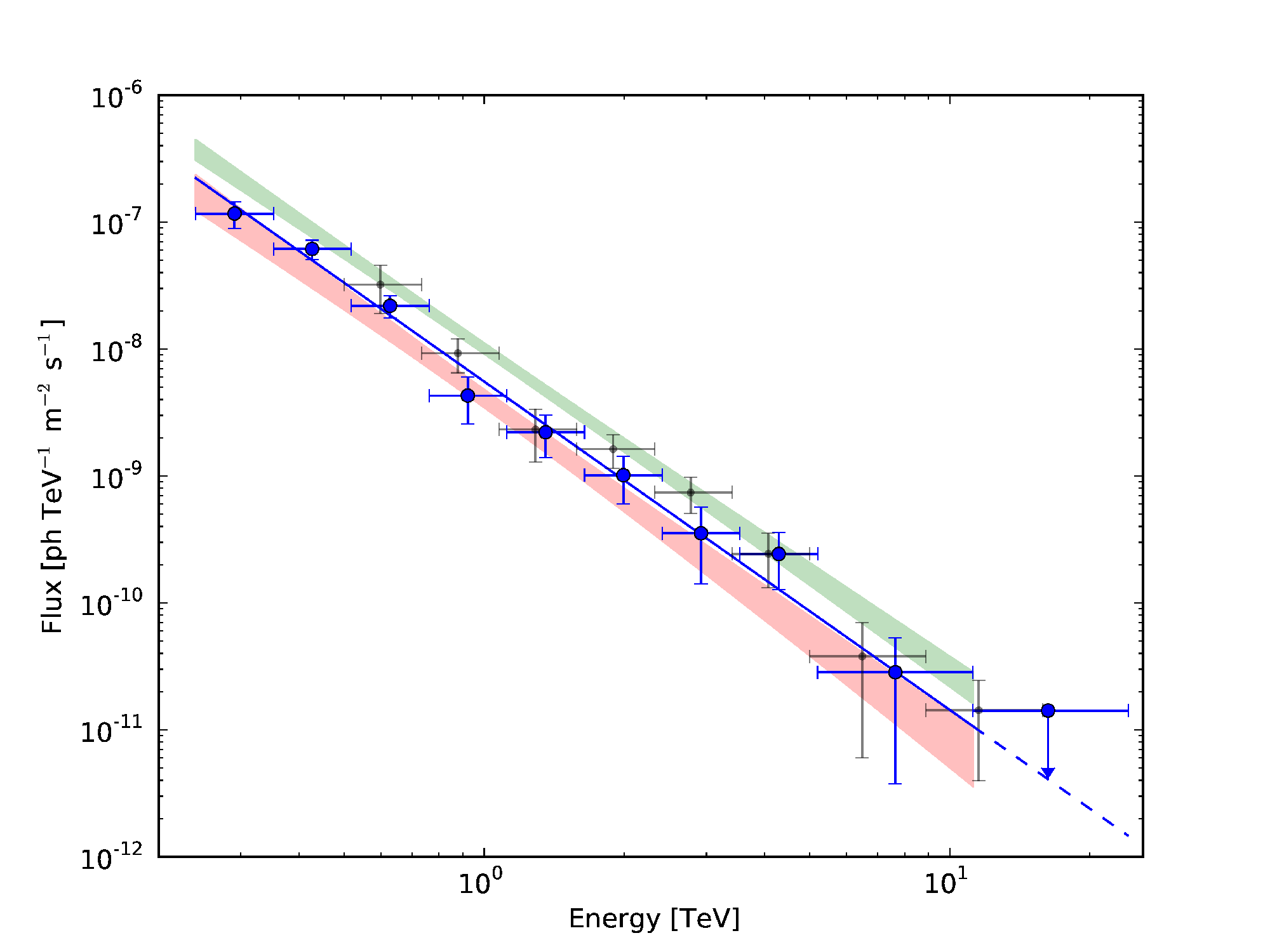}
\caption{\label{fig:vtsspectrum}The measured VHE spectrum from
  1ES\,0229+200 averaged over all three seasons (blue points with
  error bars). The upper (green) and lower (red) shaded regions show
  the spectral shape during the flaring and low periods,
  respectively. The black points are the archival H.E.S.S. spectral
  points from \citet{Aharonian:2007dn}.}
\end{figure}

The spectrum shown in Figure \ref{fig:vtsspectrum} can be fitted with
a simple power law, and the resulting normalization (at 1 TeV) and
photon index are $(5.54 \pm 0.56_{\rm stat} \pm 1.10_{\rm sys}) \times
10^{-9}$ m$^{-2}$ TeV$^{-1}$ s$^{-1}$ and $2.59 \pm 0.12_{\rm stat}
\pm 0.26_{\rm sys}$ respectively, with a $\chi^2$ of 5.8 with 7
degrees of freedom \citep[the spectral analysis procedure is described
  in][]{Acciari:2008oq}. The systematic errors on the normalization
and index based on observations of the Crab Nebula are 20\% and 10\%,
respectively. These results are comparable with those seen by
H.E.S.S., confirming the previously measured hardness.  The data for
the spectral points are given in Table \ref{tab:vtsspectrum}.  The
spectral shapes were also derived individually for the first observing
period (MJD 55118 - 55131), when the flux was high, and for the
remaining low periods (MJD 55144 - 55951).  These are shown as shaded
regions in Figure \ref{fig:vtsspectrum}. No significant change is
observed in the photon index. The index is $2.53 \pm 0.11_{\rm stat}
\pm 0.25_{\rm sys}$ during the high period and $2.64 \pm 0.19_{\rm
  stat} \pm 0.26_{\rm sys}$ during the low period while the
normalization rose from $(4.13 \pm 0.65_{\rm stat} \pm 0.83_{\rm
  sys})\times 10^{-9}$ m$^{-2}$~TeV$^{-1}$~s$^{-1}$ to $(10.2 \pm
1.0_{\rm stat} \pm 2.0_{\rm sys})\times 10^{-9}$
m$^{-2}$~TeV$^{-1}$~s$^{-1}$.

\section{{\it Swift} Observations and Results}

\begin{table}
\begin{center}
  \caption{List of {\it Swift} observations during the first season of
    observations with VERITAS.  There is an additional 3\% systematic
    error on the XRT index and XRT flux.\label{tab:swift}}
  \smallskip
\begin{tabular}{ccccccc}
  \tableline\tableline
  Obs. ID. & Date & XRT Exp. & XRT Ind. & XRT Flux & Cstat/dof & UVOT\\
  & [MJD] & [s] & & [$10^{-11} {\rm erg~cm}^{-2} {\rm s}^{-1}$] & & Filter\\ 
  \tableline
31249004 & 55117.305 & 969 & $2.07^{+0.28}_{-0.26}$ & $1.64^{+0.18}_{-0.12}$ & 132/182 & all\\
31249005 & 55118.376 & 1075 & $2.31^{+0.30}_{-0.27}$ & $1.54^{+0.12}_{-0.09}$ & 167/210 & all\\
31249006 & 55125.283 & 496 & $1.47^{+0.36}_{-0.35}$ & $1.91^{+0.30}_{-0.23}$ & 104/145 & w2 \\
31249007 & 55126.273 & 1204 & $1.58^{+0.22}_{-0.22}$ & $1.85^{+0.19}_{-0.13}$ & 181/229 & m2 \\
31249008 & 55127.346 & 1002 & $1.66^{+0.23}_{-0.22}$ & $1.91^{+0.14}_{-0.15}$ & 162/216 & w1 \\
31249009 & 55128.282 & 1020 & $1.79^{+0.26}_{-0.23}$ & $1.63^{+0.13}_{-0.13}$ & 160/206 & uu \\
31249010 & 55129.284 & 1190 & $1.82^{+0.26}_{-0.24}$ & $1.44^{+0.10}_{-0.09}$ & 141/190 & w2 \\
31249011 & 55130.288 & 1150 & $1.81^{+0.25}_{-0.24}$ & $1.56^{+0.14}_{-0.11}$ & 176/215 & m2 \\
31249012 & 55131.360 & 1170 & $1.59^{+0.21}_{-0.20}$ & $1.97^{+0.15}_{-0.16}$ & 193/240 & w1 \\
31249013 & 55152.248 & 844 & $1.89^{+0.22}_{-0.21}$ & $2.37^{+0.16}_{-0.21}$ & 178/223 & uu \\
31249014 & 55153.250 & 6770 & $1.82^{+0.18}_{-0.17}$ & $2.18^{+0.16}_{-0.11}$ & 206/267 & w2 \\
31249015 & 55154.250 & 1242 & $1.79^{+0.16}_{-0.16}$ & $2.53^{+0.19}_{-0.13}$ & 238/294 & m2 \\
31249016 & 55155.256 & 990 & $1.98^{+0.18}_{-0.18}$ & $2.46^{+0.15}_{-0.14}$ & 168/254 & w \\
31249017 & 55156.259 & 826 & $1.92^{+0.23}_{-0.21}$ & $2.41^{+0.15}_{-0.15}$ & 178/228 & uu \\
31249018 & 55157.262 & 1018 & $1.79^{+0.17}_{-0.17}$ & $2.62^{+0.17}_{-0.14}$ & 193/276 & w2 \\
31249019 & 55158.267 & 952 & $1.89^{+0.23}_{-0.21}$ & $2.00^{+0.11}_{-0.17}$ & 183/228 & w2 \\
  \tableline
average & N/A & 21920 & $1.68^{+0.05}_{-0.05}$ & $2.30^{+0.03}_{-0.04}$ & 617/649 & mult.\\
  \tableline
\end{tabular}
\end{center}
\end{table}

The {\it Swift} data set contains sixteen snapshot observations
ranging from 505 to 1394 seconds in duration as shown in Table
\ref{tab:swift}.  All {\it Swift} X-ray Telescope
\citep[XRT,][]{Burrows:2005yg} data are reduced using the HEAsoft 6.13
package\footnote{\label{note:heasoft}https://heasarc.gsfc.nasa.gov/lheasoft/}. Event
files are calibrated and cleaned following the standard filtering
criteria using the xrtpipeline task and applying the most recent {\it
  Swift}-XRT calibration files (Update 2012-02-09). All data were
taken in window timing (WT) mode and no pile-up is seen.  Rectangular
source and background regions were used with a length of 40 pixels
along the data stream.  The XRT data were fitted with an absorbed
power law using the absorption calculated by
\citet{Kaufmann:2011fk}($N_{\rm H}$ = $1.1 \times 10^{21}~{\rm
  cm}^{-2}$). We did not test for curvature and the goodness of fit
was evaluated using the C-statistic (shown in Table \ref{tab:swift}).
The hard spectrum (photon index $\sim$ 1.7) and UV (see below) to
X-ray SED suggest that the synchrotron emission peaks above 10 keV
(see Figure \ref{fig:SyncPeak}). Table \ref{tab:swift} shows the flux
and photon index measured by the XRT in the 0.2 - 10 keV energy
range. Note that there is an additional 3\% systematic error on the
XRT flux and XRT
index.\footnote{http://www.swift.ac.uk/analysis/xrt/files/SWIFT-XRT-CALDB-09\_v16.pdf}
We fit a constant line to both the flux and index data and the
$\chi^2$ for the fits to constant flux and constant index are 126 and
9.1 with 15 degrees of freedom, respectively. Thus, the {\it
  Swift}-XRT flux is variable at a level of 9.1 standard deviations
and the $\chi^2$ probability of the photon index being constant is
87.3\%. Note that this probability is quite high which indicates that
the uncertainties might be overestimated.  The normalized excess
variance \citep{Vaughan:2003fk}, which is an indicator of the
underlying variability of the source taking into account statistical
errors, of the XRT data is 0.063 $\pm$ 0.013 corresponding
to a fractional variability of 25\%.  The doubling time based on
the change in flux between the first XRT observing period (MJD 55117 -
55131) and the second (MJD 55152 - MJD 55158) is $73 \pm 16$ days.
However, the doubling time from the lowest flux state which occurred
on MJD 55129 and the highest flux on MJD 55157 is $33 \pm 8$ days.

{\it Swift} Ultraviolet and Optical Telescope
\citep[UVOT,][]{Poole:2008do} observations were taken in several
different photometric bands since the choice of filter was left to the
discretion of the {\it Swift} operations team (see Table
\ref{tab:swift} for more details). The {\it uvotsource} tool is used
to extract counts, correct for coincidence losses, apply background
subtraction, and calculate the source flux.  The UVOT data were
corrected for interstellar extinction \citep{Seaton:1979kx} and dust
absorption \citep{Schlegel:1998lw}.  There is still substantial
host-galaxy contamination, especially in the B and V bands, which was
corrected using the correction factors derived by
\citet{Kaufmann:2011fk}\footnote{Since \citeauthor{Kaufmann:2011fk}
  did not estimate the host-galaxy contamination in the UVW1 and UVM2
  filters, we assume that it can be comprised between 0 and 30\%
  (which is the value computed for the nearby U filter). In both
  Figures 5 and 6, the error bars for the filters UVW1 and UVM2
  include this systematic uncertainty.}.  Figure \ref{fig:SyncPeak}
presents both the corrected and uncorrected UVOT data for comparison.
The average flux measured by UVOT (in units of $10^{-16}\,\rm
erg\,cm^{-2}s^{-1}\AA^{-1}$) is $F_B = 3.09 \pm 0.25$, $F_V = 4.77 \pm
0.38$, $F_U = 3.98 \pm 0.12$, $F_{M2} = 6.21 \pm 0.28$, $F_{W2} = 6.74
\pm 0.28$, $F_{W1} = 5.33 \pm 0.21$ (the reported errors include
statistical and systematic errors summed in quadrature).  The
normalized excess variance in each band is a small negative number
ranging from -0.07 to -0.005 indicating that any inherent variability
is within the measurement errors. The UVOT data are shown in the
lightcurve (Figure \ref{fig:fulllightcurve}) and in the SED figure
(Figure \ref{fig:SED}).

\begin{figure}
\plotone{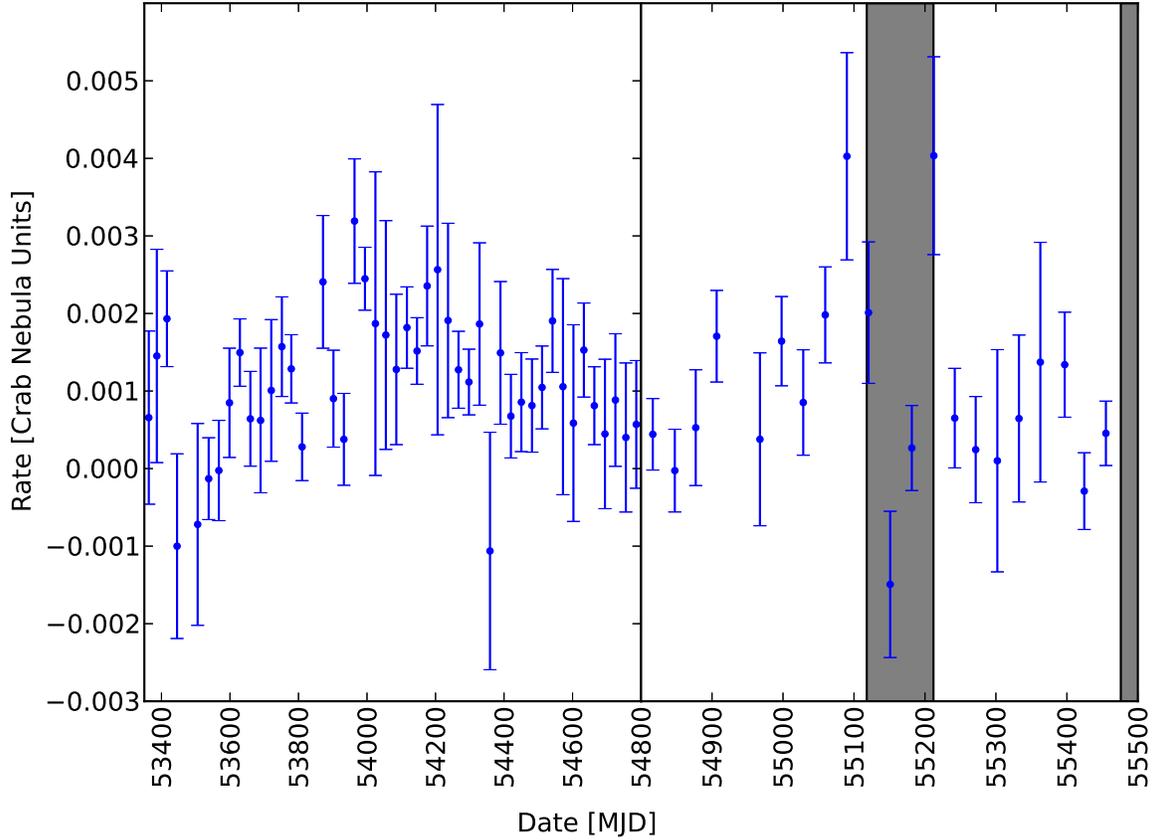}
\caption{\label{fig:batlightcurve}Rate in the BAT 70-month survey (in
  units of the Crab Nebula's flux) from 14 to 195 keV. The abscissa of
  the right panel is spread out for clarity but the time series
  between the two panels is continuous. Since the BAT is a survey
  instrument, the exposure on 1ES\,0229+200 is not constant in each
  bin and this is reflected in the size of the error bars.  Two points
  with very small exposures (686 and 141 seconds respectively) and
  thus very large error bars at MJD 53841 and 54937 have been removed
  for clarity. The first grey band is the extent of the VERITAS
  2009-2010 observing season listed in Table \ref{tab:vtsobs} and
  shown in Figure \ref{fig:vtslightcurve}.  The VERITAS 2010-2011
  observing season begins at the end of the BAT 70-month survey and
  can be seen as a grey band on the far right.  The BAT 70-month
  survey provides data up to September 2010.  These data are also
  shown in context with the other data sets in Figure
  \ref{fig:fulllightcurve}.}
\end{figure}

1ES\,0229+200 is also in the {\it Swift}-BAT 70-month hard X-ray
survey \citep{Baumgartner:2012ly} which includes data from December
2004 to September 2010.  This survey contains sources detected in the
BAT in the 14 - 195 keV band down to a significance level of 4.8
$\sigma$.  The blazar is detected at a flux level of
$24.50^{+4.54}_{-4.23}\times10^{-12}~{\rm erg~s}^{-1}~{\rm cm}^{-2}$
with a spectral index of $2.16^{+0.28}_{-0.25}$ ($\chi^2_{\rm r} =
0.70$). Both statistical and systematic errors are included in the
quoted errors above. The overall SED from the BAT is shown in Figures
\ref{fig:SyncPeak} and \ref{fig:SED} and the 70-month light curve is
shown in Figure \ref{fig:batlightcurve}. The normalized excess
variance of the BAT light curve is $0.48 \pm 4.50$ corresponding to a
fractional variability of 69\%.

\section{RXTE Observations and Results}

\begin{figure}
\epsscale{0.6} 
\plotone{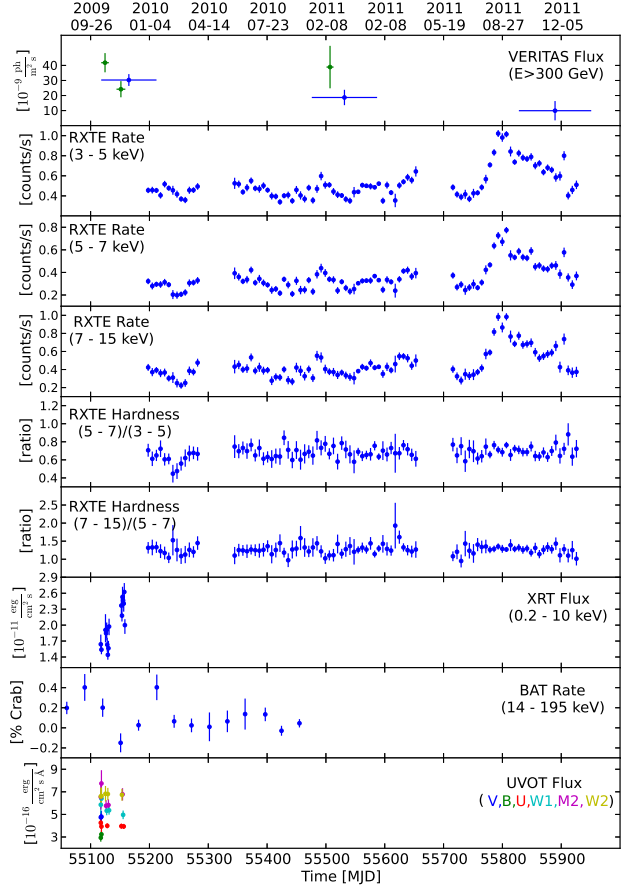}
\caption{\label{fig:fulllightcurve} The upper panel shows the VHE flux
  measured by VERITAS and detailed in Figure
  \ref{fig:vtslightcurve}. The green points are the data binned by
  observing period.  Only points with a significance greater than two
  standard deviations are shown (the full data set can be found in
  Table \ref{tab:vtsobs}). The second through fourth panels show low
  (3 - 5 keV), mid (5 - 7 keV), and high (7 - 15 keV) energy range
  count-rate light curves from the RXTE PCA binned by week.  The fifth
  and sixth panels show the hardness ratio from the RXTE PCA (mid
  energy to low energy bands and high energy to mid energy band,
  respectively). The seventh panel shows the flux measured by XRT in
  the 0.2 - 10 keV band and the eighth panel shows the rate in
  the BAT in the 14 - 195 keV band.  The last panel shows the flux
  measured using the six different UVOT filters (V is blue, B is
  green, U is red, UVW1 is cyan, UVM2 is magenta and UVW2 is yellow).}
\end{figure}

1ES\,0229+200 was the target of several RXTE \citep{Bradt:1993fk}
monitoring campaigns during the VERITAS observations.  This resulted
in robust coverage between 2 and 20 keV.  The RXTE Proportional
Counter Array (PCA) data were reduced using the HEAsoft 6.13 package
and the most recent background models\footnote{August 6, 2006 release
  from http://heasarc.gsfc.nasa.gov/docs/xte/pca\_news.html}.  Only
data from layer 1 of PCU 2 were used, in order to maximize the signal
to noise ratio.  The
suggested\footnote{http://heasarc.gsfc.nasa.gov/docs/xte/}
conservative cuts on the Earth observation angle, pointing offset, SAA
passage time, and electron contamination were used to filter the
data. The spectrum in each individual campaign was fit with an
absorbed ($N_{\rm H}$ = $1.1 \times 10^{21}~{\rm cm}^{-2}$) power law
resulting in fluxes of $2.67^{+0.02}_{-0.02}, 2.94^{+0.07}_{-0.06}$
and $3.75^{+0.02}_{-0.04} \times 10^{-11}~{\rm erg~cm}^{-2} {\rm
  s}^{-1}$ and photon indices of $1.92^{+0.04}_{-0.04},
1.90^{+0.10}_{-0.09}$ and $1.87^{+0.03}_{-0.02}$ with $\chi^2$/dof of
32/26, 13/26 and 29/26 respectively.  The three campaigns covered the
time periods MJD 55198 to 55282, MJD 55345 to 55653 and MJD 55715 to
55926. Figure \ref{fig:SyncPeak} shows the RXTE spectra and Figure
\ref{fig:fulllightcurve} shows the count-rate light curve from the PCA
in three different wavebands (the three campaigns can clearly be
seen).  There is significant variability in these data, including a
large flare before the 2011-2012 VERITAS observing season.  Hardness
ratios are also plotted in Figure \ref{fig:fulllightcurve}, but no
significant change in the spectral shape is seen during the three
years (this is confirmed by the spectral analysis above). The
normalized excess variance (fractional variability) of the RXTE data
is $0.050 \pm 0.021~(22\%)$, $0.037 \pm 0.016~(19\%)$ and $0.043 \pm
0.014~(21\%)$ for the high (7 - 15 keV), mid (5 - 7 keV) and low (3 -
5 keV) bands respectively.

\section{Discussion}

\subsection{Temporal Analysis}

\begin{figure}
\plotone{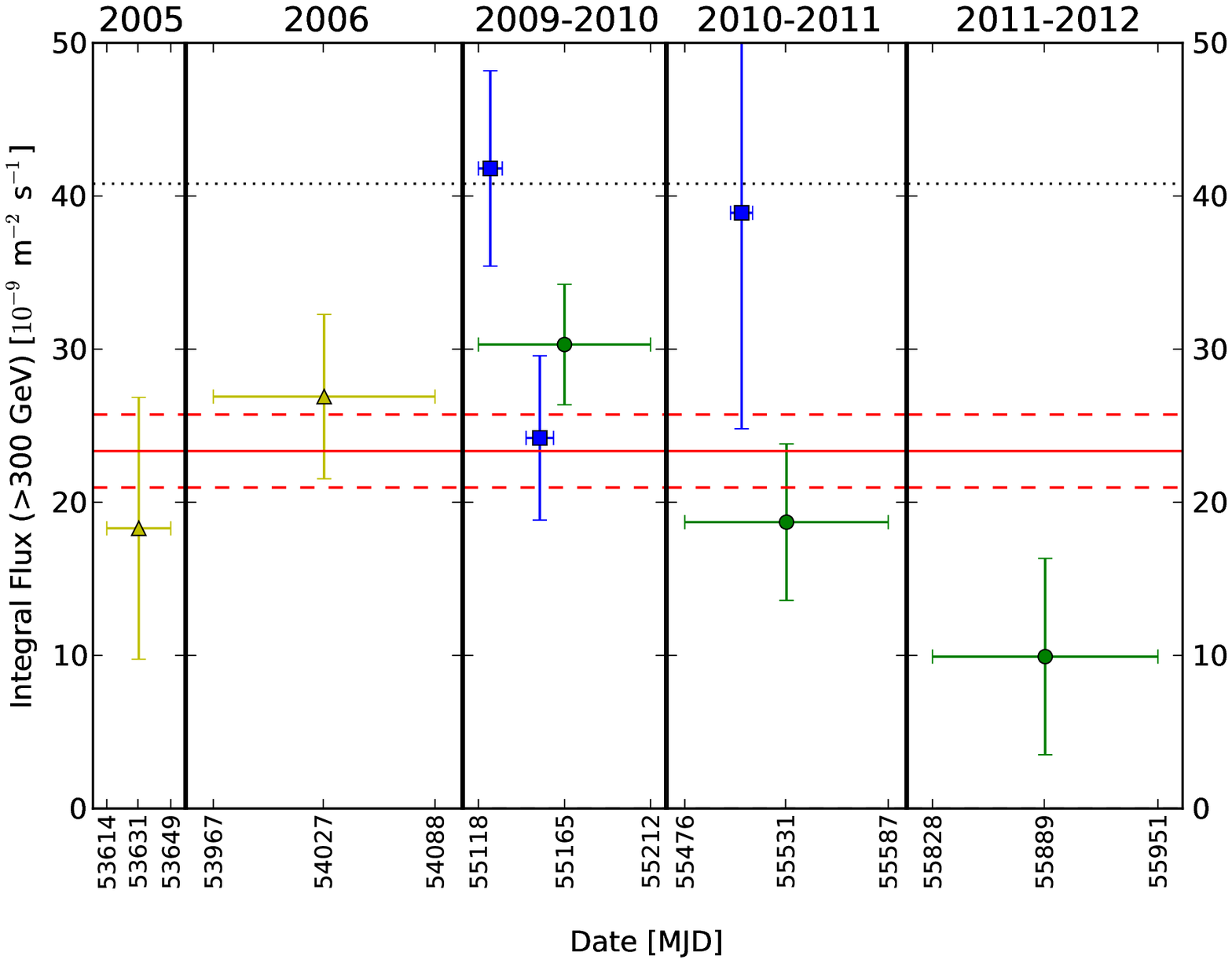}
\caption{\label{fig:vtslightcurve}Integral flux above 300 GeV for
  1ES\,0229+200, binned by observing season (green circles). The
  vertical black lines also delineate the observing seasons.  The
  yellow triangles in 2005 and 2006 are from the previous
  H.E.S.S. measurements \citep{Aharonian:2007dn}, shown for
  comparison.  The blue squares are the data binned by observing
  period. Only points with a significance greater than two standard
  deviations are shown (the full data set can be found in Table
  \ref{tab:vtsobs}).  The horizontal red lines (solid is the value and
  dashed is the statistical error range) are the fit to the VERITAS
  yearly data (green circles).  The black dotted line shows a 3\% Crab
  Nebula flux for comparison.  These data are also shown in context
  with the other data sets in Figure \ref{fig:fulllightcurve}.}
\end{figure}

The full VHE light curve is shown in Figure \ref{fig:vtslightcurve}. A
fit of a constant flux to the VERITAS yearly points (shown as the
solid red line in the figure, $(22.9 \pm 2.8) \times 10^{-9}$ m$^{-2}$
s$^{-1}$) yields a $\chi^2$ of 8.32 with 2 degrees of freedom
(probability of 1.6\%).  The blue squares in Figure
\ref{fig:vtslightcurve} show the data divided into individual
observing periods, for those with statistical significance above 2
sigma (all points are given in Table \ref{tab:vtsobs}). The normalized
excess variance is $3.9 \pm 1.7$ for the yearly binned data and $0.038
\pm 0.038$ for the data binned by observing period.  This corresponds
to a fractional variability of 200\% for the yearly lightcurve and
19\% for the monthly lightcurve. A fit of a constant flux to the
monthly data (including all observing periods, as listed in Table 1)
results in a $\chi^2$ value of 24.7 with 13 degrees of freedom or a
probability of being constant of 2.5\%. The evidence for variability
in these data is not conclusive by itself, but, when considered in the
context of a known variable source class and significant variability
in the X-ray band, we consider it to be indicative of truly variable
emission.

The X-ray flux as measured by {\it Swift}-XRT is variable at a level
of 9.1 standard deviations and the PCA data shown in Figure
\ref{fig:fulllightcurve} show evidence for variability throughout the
three seasons (fractional variability $\sim$ 20\%), including a large
flare preceding the 2011-2012 VERITAS observing season. A constant
flux in the RXTE data is excluded at greater than ten standard
deviations in all three bands.  The hard X-ray data from the BAT shown
in Figure \ref{fig:batlightcurve} display an interesting feature.
Directly preceding the first season of VERITAS observations (where the
highest VHE fluxes were measured), the BAT flux from 1ES\,0229+200
reached a level not previously seen in the lifetime of the BAT
instrument.  The BAT flux then dropped to one of the lowest levels
seen.  The high flux was repeated at the end of the VERITAS observing
season, where observations at VHE were not possible due to the full
moon.  If the same particle population is involved in both the high-
and low-frequency emission, the X-ray variability seen in RXTE and
{\it Swift} implies that variability should be seen at VHE.

Most previous studies using distant blazars to place a lower limit on
the IGMF must assume that the measured VHE spectrum (exposure time on
the order of tens of hours) is a good estimator of the time-averaged
spectrum of the source over several years or more. Since the flux from
1ES\,0229+200, and several other VHE blazars had previously been
consistent with a constant-emission model, this assumption was made by
some authors attempting to limit the strength of the IGMF
\citep{Arlen:2012kx, Dermer:2011fk, Huan:2011vn, Neronov:2010uq,
  Georganopoulos:2010ys}.  However, for variable sources, the
multi-year time-averaged differential flux is unknown and difficult to
estimate with any reliability. Because of this inherent ambiguity, any
lower limit on the IGMF derived using the measured VHE spectrum from
variable sources is not robust. The observations presented here show
that the constant-flux hypothesis may not be valid for 1ES 0229+200,
as shown in Figure \ref{fig:vtslightcurve} and Table \ref{tab:vtsobs}.
Based on archival observations of 1ES\,0229+200, \citet{Dermer:2011fk}
determined that $B_{IGMF} \gtrsim 5\times 10^{-18}~{\rm G}$ assuming a
variability time scale of $\sim3$ years.  The observations presented
here show that 1ES\,0229+200 is variable on at least a yearly
timescale.  Since the spectral shape is not changing and the derived
limit scales as the square root of the time scale, we can assume that
this reduces the limit by a factor of $\sqrt{1/3}$ to $B_{IGMF}
\gtrsim 3 \times 10^{-18}$~G.  The detection of variability also
modifies the conclusions of \citet{Arlen:2012kx} (who assumed that the
original HESS measurement was characteristic of the average flux).
They ruled out a zero IGMF hypothesis (H0) at 99\% confidence based
upon the 1ES\,0229+200 spectrum but stated that if variability is
detected in this blazar the H0 hypothesis is not ruled out at more
than 95\% confidence (1ES\,0229+200 is the only blazar in their sample
that could have rejected the H0 hypothesis).

Several authors have developed quantitative models of the plasma
physics of pair cascades in the IGMF. \citet{Broderick:2012uq} and
\citet{Schlickeiser:2012ys} place doubts on whether measurements of
blazar spectra can be used to place constraints on the IGMF even if
the source is non-variable since plasma instability losses dominate
over inverse Compton loses.  However \citet{Broderick:2012uq} concedes
that, during a flare, IC losses might dominate since instabilities
have not promptly set in. \citet{Miniati:2013zr} contradict this by
stating that the relaxation time of a plasma beam is much longer than
the IC cooling time so that the beam can be stable and allow for
secondary $\gamma$-ray emission.  In conclusion, even if 1ES\,0229+200
is non-variable, it is debatable that a meaningful constraint on
the magnitude of the IGMF could be derived depending on the exact
physics of the pair cascade plasma beam.

\subsection{Spectral Analysis}

\begin{figure}
\plotone{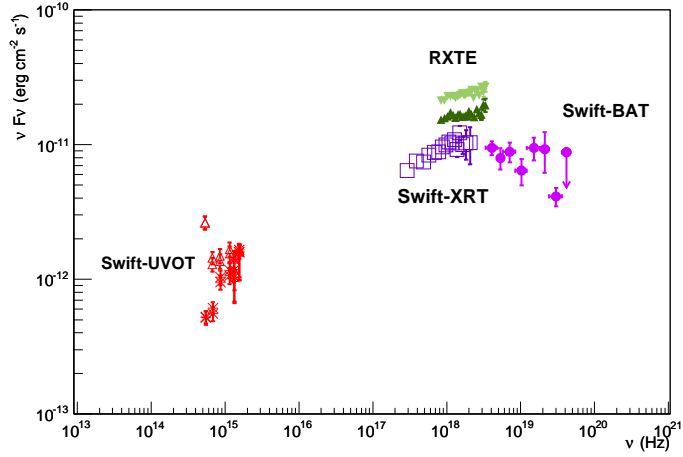}
\caption{\label{fig:SyncPeak}The synchrotron part of the SED of the
  blazar 1ES\,0229+200.  The red triangles are the uncorrected {\it
    Swift}-UVOT measurements while the red stars are the absorption and
  host-galaxy corrected UVOT points.  The synchrotron peak is
  constrained by the {\it Swift}-XRT, {\it Swift}-BAT and the RXTE
  measurements in the X-ray.  Two RXTE periods are shown, a low state
  and during the large flare around MJD 55800.}
\end{figure}

\begin{figure}
\plotone{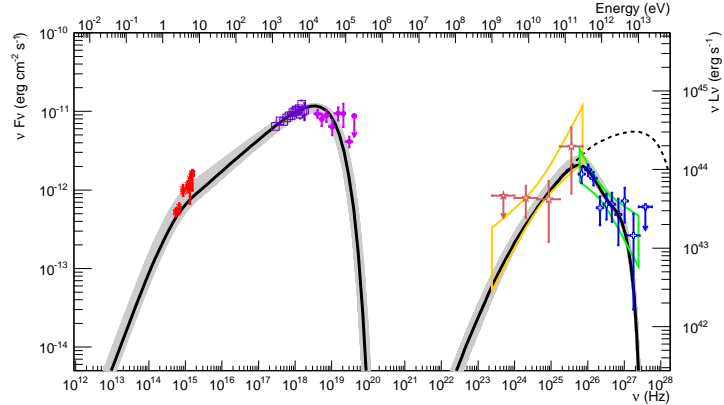}
\caption{\label{fig:SED}The multi-wavelength SED of the blazar
  1ES\,0229+200.  The red asterisks are the average {\it Swift}-UVOT
  measurements, the open purple squares are the average {\it
    Swift}-XRT measurements and the pink circles are the 70-month
  average {\it Swift}-BAT measurements.  The BAT and XRT data straddle
  the synchrotron peak.  At high energies, the {\it Fermi}-LAT points
  are shown as salmon stars (the yellow region indicates the
  statistical uncertainty on the spectral model) and the VERITAS data
  are shown as blue crosses (the green region indicates statistical
  uncertainty on the spectral model).  The UVOT data are corrected for
  absorption and for host-galaxy emission (see the text).  The grey
  region indicates the range of all of the one-zone SSC models which
  correctly describe the SED. The solid black curve is the SSC model
  with the lowest $\chi^2$ value with respect to the data. The dotted
  black line represents the best-fit SSC model before absorption on
  the EBL.}
\end{figure}

\begin{figure}
\centering
\begin{minipage}{.5\textwidth}
\centering
\includegraphics[scale=0.4]{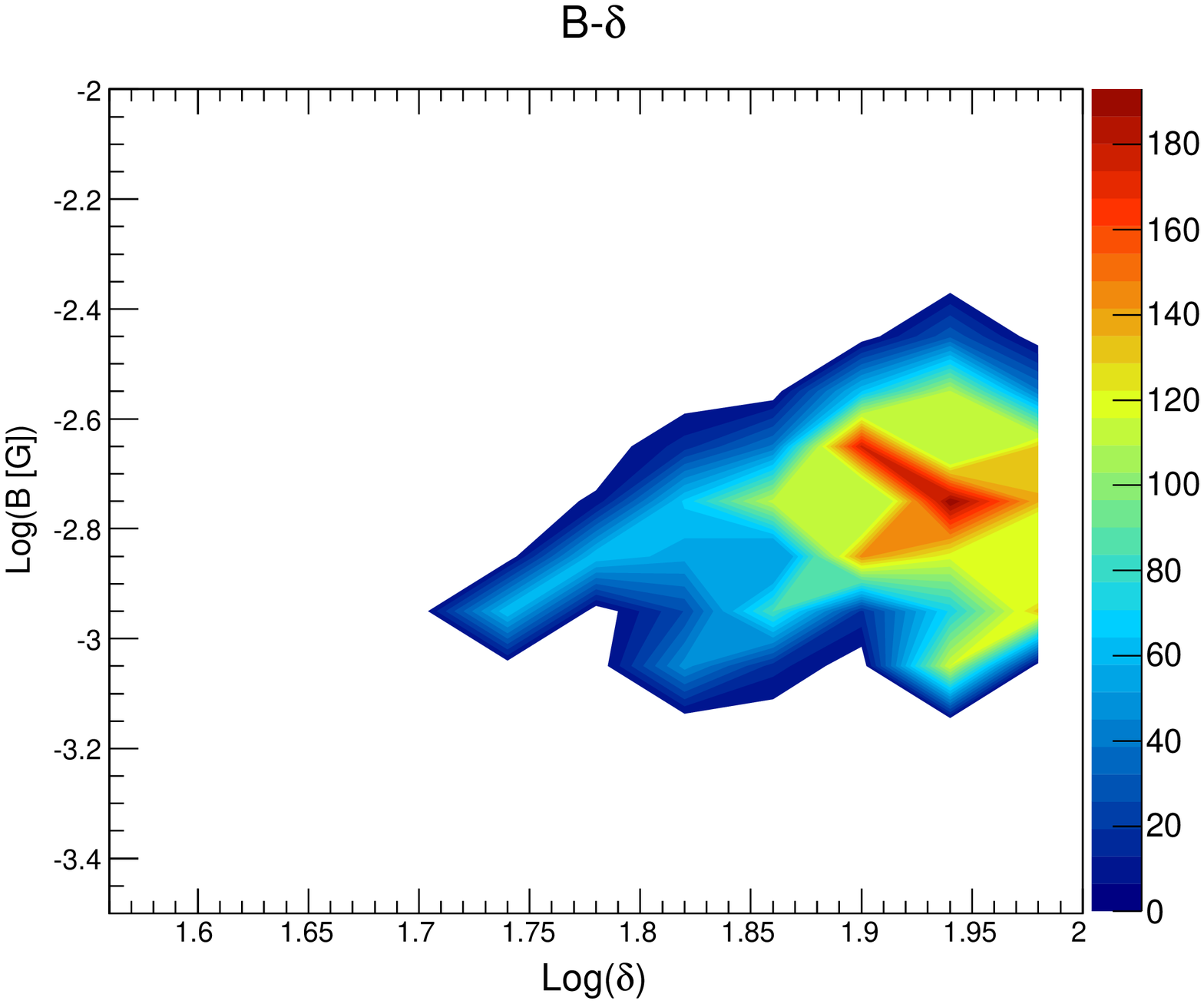}
\end{minipage}%
\begin{minipage}{.5\textwidth}
\centering
\includegraphics[scale=0.4]{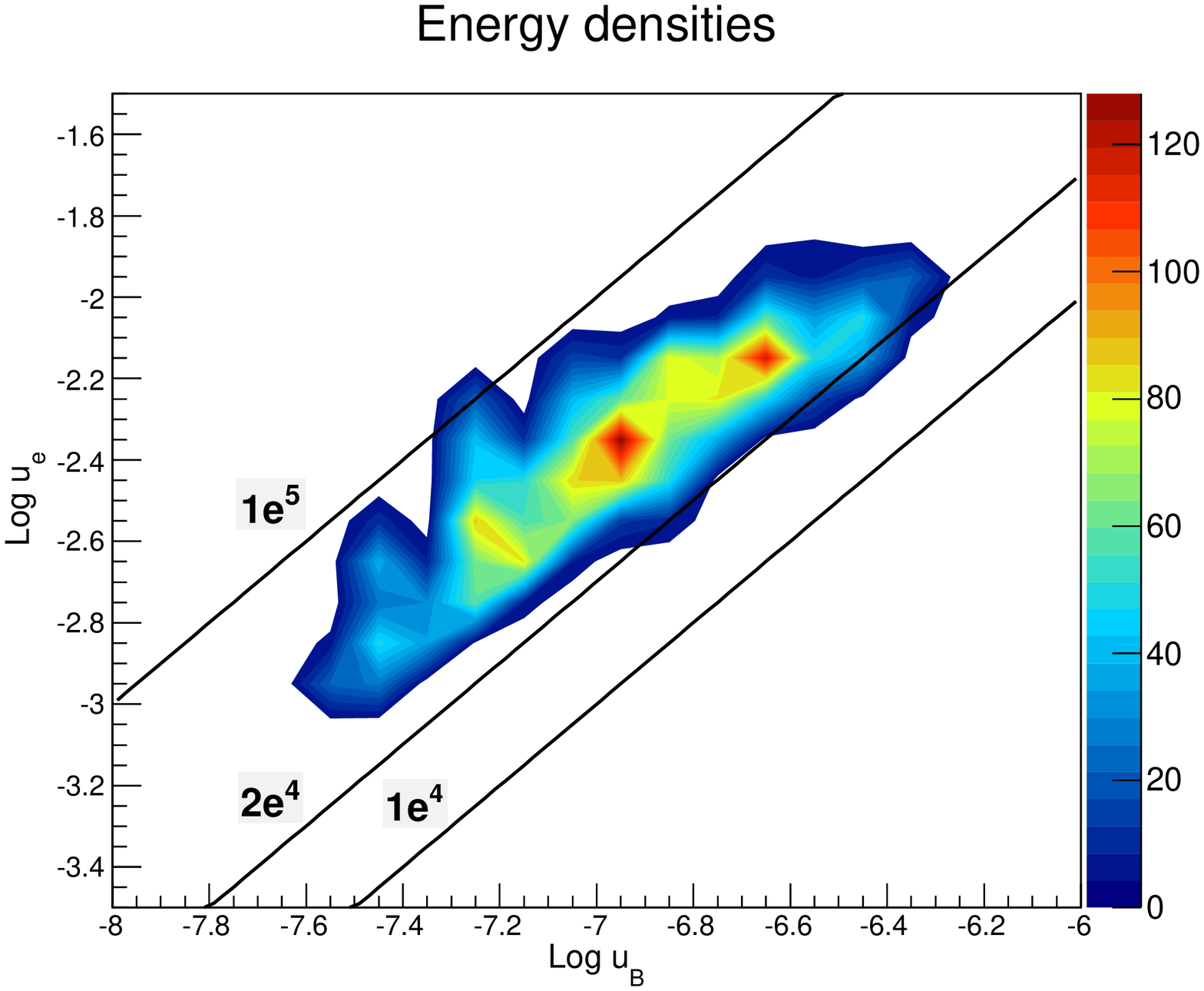}
\end{minipage}
\caption{\label{fig:Contours} Left: $\log_{10}{B}$-$\log_{10}{\delta}$
  parameter space for the SED shown in Figure \ref{fig:SED}.  The
  color scale is arbitrary and the most extended contour represents
  the 1 $\sigma$ region.  Values of $\delta$ higher than $100$ have
  not been studied. Right: $\log_{10}{U_e}$-$\log_{10}{U_B}$ SED
  parameter space.  The color scale is arbitrary\ and the most
  extended contour represents the 1 $\sigma$ region.  The slanted
  lines are equipartition contours ($U_e/U_B$).}
\end{figure}

The broadband SED of 1ES\,0229+200 is shown in Figure \ref{fig:SED}
and the synchrotron peak is detailed in Figure \ref{fig:SyncPeak}. The
{\it Fermi}-LAT data, as well as the best-fit LAT bow-tie are
reproduced from \citet{Vovk:2012vn}, where they reported a detection
at the level of seven standard deviations in almost three years of
observations.  \citet{Vovk:2012vn} fitted the high energy spectrum
from 1 to 300 GeV using a power law with a spectral index of $\Gamma =
1.36 \pm 0.25$ and normalization at 20 GeV of $N_0 = (1.4 \pm 0.5)
\times 10^{-15}$ MeV cm$^{-2}$ s$^{-1}$.  Note that the {\it
  Swift}-BAT, {\it Fermi}-LAT, and VERITAS data are long-term average
spectra (70 months for the BAT, 3 years for the LAT, and VERITAS),
while the {\it Swift}-XRT and UVOT data are short-term averages taken
during the initial VERITAS observing season in 2009-2010. The UVOT
spectral points plotted in Figure \ref{fig:SED} have been corrected
for host-galaxy contribution, using the correction factors from
\citet{Kaufmann:2011fk}. The availability of {\it Swift}, {\it Fermi},
and VERITAS data allows both the low-energy and high-energy peaks to
be constrained. The {\it Swift}-XRT spectrum is especially hard,
indicating that the synchrotron peak is located above the XRT energy
band (see Figure \ref{fig:SyncPeak}), but the additional information
from the BAT suggests that the peak is located between the two bands
at $E \simeq 10$ keV.

The SED is modeled using the one-zone synchrotron-self-Compton (SSC)
code of \citet{Katarzynski:2001vn}, taking into account EBL
attenuation based upon the calculations of
\citet{Franceschini:2008wo}\footnote{The
  \citeauthor{Franceschini:2008wo} EBL template used here is in
  agreement with the EBL measurements of both the {\it Fermi}-LAT
  \citep{Ackermann:2012ly} and H.E.S.S. collaborations
  \citep{Abramowski:2013qf}.}  The SSC parameter space is constrained
by the algorithm described in \cite{Cerruti:2013zr}, which can be seen
as a numerical extension of the constraints defined by
\citet{Tavecchio:1998fk}, using in addition the information from {\it
  Fermi}-LAT and VERITAS. The basic idea is to define a system of
equations linking SSC parameters and physical observables, and to
solve this system in order to determine the set of SSC parameters
which correctly describes the SED. The system of equations is obtained
numerically, simulating a grid of SSC models, determining for each of
them the expected values of the observables, and then performing a fit
to find the best parameterization of each observable as a function of
the model parameters. Given the uncertainty in the physical
observables, the system of equations is solved iteratively, spanning
each observable in the range $\pm 1\sigma$.

The SSC model by \citet{Katarzynski:2001vn} assumes a spherical
emission volume of radius $R$ moving towards the observer with Doppler
factor $\delta$, and filled with a tangled, homogeneous magnetic field
$B$ and a nonthermal population of electrons and/or positrons
$N_e(\gamma_{\rm e})$. The particle distribution is parameterized by a
single power-law function (with index $\alpha$ and normalization $K$),
between minimum and maximum Lorentz factors $\gamma_{\rm min}$ and
$\gamma_{\rm max}$, respectively.  Note that the modeling presented
here ignores any possible contribution to the SED from photons
reprocessed by the IGMF.

Thus, the SSC model has seven free parameters: $\delta$, $B$ and $R$,
for the emitting region, and $\alpha$, $\gamma_{\rm min;max}$ and $K$,
for the particle distribution, where $K$ is defined as the electron
number density at $\gamma_{\rm max}$. In order to determine the set of
solutions which correctly describe the SED, we used seven physical
observables: the synchrotron peak frequency and flux, the X-ray
spectral index, the {\it Fermi} and VERITAS fluxes at their respective
decorrelation energies, and the {\it Fermi} and VERITAS spectral
indices.

Following \cite{Cerruti:2013zr}, we then simulated SSC models within
the following parameter space: $\delta \in [40,100]$, $B \in [0.0005,
  0.01]~\textrm{G}$, $R \in [10^{15}, 10^{17}]~\textrm{cm}$,
$\gamma_{\rm min} \in [10^{4}, 10^{5}]$, $\gamma_{\rm max} \in
[10^{6}, 2\times10^{7}]$, and $K \in [10^{-12},
  10^{-9}]~\textrm{cm}^{-3}$. The value of $\alpha$ is fully
constrained by the value of the {\it Swift}-XRT spectral index: to
take into account its uncertainty we computed three different sets of
solutions, for $\alpha=2.18$, $2.24$, and $2.30$. For each simulated
SSC model we determined the values of the observables, producing a
grid containing, for each combination of the six parameters, the
corresponding values of the six observables. It is important to note
that, when computing the simulated spectral index observed in the
VERITAS energy range, we excluded the upper limit at $E \simeq 16$
TeV. In fact, given the strong EBL absorption at $E \geq 10$ TeV, the
simulated SSC models significantly under-estimate the VERITAS
power-law fit at $E \simeq 16$ TeV, even though they are fully
consistent with the flux upper limit.

The grid is then fitted, determining the system of equations linking
parameters and observables, which is then solved for the specific case
of 1ES\,0229+200. We iteratively solved the system spanning the range
(in logarithm, except for the {\it Fermi} and VERITAS spectral
indices): $\nu_{\rm sync-peak} \in [18.45,18.55]$ Hz ([11.66,14.67]
keV), $\nu F_{\nu; {\rm sync-peak}} \in [-10.99,-10.93]$ erg cm$^{-2}$
s$^{-1}$, $\Gamma_{Fermi} \in [1.08, 1.64]$, $\Gamma_{\rm VERITAS} \in
[2.30, 2.88]$, $\nu F_{\nu; Fermi} \in [-12.25,-11.91]$ erg cm$^{-2}$
s$^{-1}$ and $\nu F_{\nu; \rm VERITAS} \in [-12.15,-11.96]$ erg
cm$^{-2}$ s$^{-1}$. For the $\gamma$-ray observables, the uncertainty
includes systematic errors, summed in quadrature with the statistical
errors. The system of equations includes an inequality relating the
variability time-scale (fixed to 33 days, corresponding to the
doubling time-scale measured in the {\it Swift} data) to the size and
the Doppler factor of the emitting region, and we reject solutions
with $\delta>100$ which would represent a strong violation of the
constraints determined from radio observations of relativistic jets
\citep[see, for example,][]{Lister:2013} or from the unification model
of AGN \citep{Henri:2006}. In order to reproduce the break observed by
UVOT, we introduced in the algorithm a new observable, the break
frequency $\nu_{break}$, defined as the intersection of the two
power-law functions fitted between $10^{10}$ and $10^{12}$ Hz, and
$10^{16}$ and $10^{17}$ Hz. We then computed a new equation linking
$\nu_{break}$ to the six free parameters. Once the set of solutions is
obtained, we then select only those solutions which are characterized
by $\log_{10}\nu_{break} \in [14.4, 14.6]$ Hz.

To verify the accuracy of our result we produce an SSC model for each
solution and compute the $\chi^2$ with respect to the observational
data (XRT, LAT and VERITAS only) in order to determine the solution
that minimizes the $\chi^2$ and to check that all the solutions are
included in a one sigma confidence interval.

The best-fit parameter values are given in Table \ref{tab:SED},
alongside two previous modeling efforts by \citet{Tavecchio:2009ys}
and \citet{Kaufmann:2011fk}. Figure \ref{fig:Contours} details the
possible values for B and $\delta$. The main difference with respect
to these previous models comes from the value of the synchrotron peak
frequency, which, in our SED, is located between the XRT and the BAT
energy bands, at $\nu_{sync-peak} \simeq 3\times 10^{18}$ Hz (12.4
keV), an order of magnitude less than that reported by
\citet{Kaufmann:2011fk} ($3.5\times 10^{19}$ Hz), but more in line
with what \citet{Tavecchio:2009ys} found ($9.1\times 10^{18}$
Hz). This difference arises mainly from the higher statistics in the
70-month BAT spectrum compared with the 58-month spectrum previously
used. Another difference compared to the previous modeling attempts of
1ES\,0229+200 in an SSC scenario is that neither
\citet{Tavecchio:2009ys} nor \citet{Kaufmann:2011fk} had the
information from the {\it Fermi}-LAT detection.

The first important result of the current modeling is that the minimum
value of the Doppler factor required to fit the SED of 1ES\,0229+200
is $\delta \geq 53$. The main observational constraint on this
parameter is the hard VHE spectral index, and the solutions
characterized by the lowest values of $\delta$ are the ones with the
softest VHE emission. This value is higher than the ones commonly
assumed in SSC modeling of HSPs \citep[see, for
  example,][]{Abdo:2011tw,Abdo:2011uq,Abramowski:2012vn} but in
agreement with the one adopted by \citet{Tavecchio:2009ys} ($\delta =
50$), while \citet{Kaufmann:2011fk} adopted $\delta = 40$. It should
be noted that $\delta = 40$ is the smallest Doppler factor that could
reproduce the spectra of 1ES\,0229+200 in \citet{Kaufmann:2011fk} and
larger values could also have been used which would have been more in
line with this work. Our solutions show some important differences
with respect to the model fit performed by \citet{Kaufmann:2011fk};
the size of the emitting region (located between $5\times
10^{15}~\textrm{cm}$ and $3 \times 10^{16}~\textrm{cm}$ or 1.62 to
9.72 mpc) is three orders of magnitude lower than that previously
derived. As a consequence, the magnetic field assumed by
\citet{Kaufmann:2011fk} is several orders of magnitude lower than the
one assumed here and in \citet{Tavecchio:2009ys}. An emitting-region
size of the order of $10^{15}$- $10^{16}~\textrm{cm}$ is similar to
the ones inferred for the VHE HSPs 1ES\,1218+304
\citep{Weidinger:2010uq} and 1ES\,1101-232 \citep{Aharonian:2007kx}.

In our modeling, we found that a parameterization of the electron
distribution by a single power-law function provides a good
description of the SED. However, a break in the spectrum is expected
in the presence of synchrotron cooling, but it is possible that the
break energy is above the value of $\gamma_{\rm max}$ (or that the
break is coincident with $\gamma_{\rm max}$, i.e., that the particle
distribution extends above $\gamma_{\rm max}$ with an index
$\alpha_2=2.3+1=3.3$). Following the study presented in
\citet{Cerruti:2013zr}, we compared our values of $\gamma_{\rm max}$
to the expectations from synchrotron cooling, and we found that
$\gamma_{\rm max}$ would be consistent with a synchrotron break only
if the injected particles are escaping from the emitting region with a
speed $v$ comprised between $c/200$ and $c/50$. If instead the
particles are escaping faster ($v\simeq c$), the synchrotron break is
expected to be at a Lorentz factor higher than $\gamma_{\rm max}$, in
agreement with our modeling.  Another interesting aspect is the energy
budget of the emitting region. For all the solutions, we compute the
values of the magnetic energy density $u_B=B^2/8\pi$ (in CGS units)
and the particle energy density $u_e=mc^2\int {d\gamma_e\gamma_e
  N_e(\gamma_e)}$. The equipartition factor $u_e/u_B$ is between
$2\times10^4$ and $10^5$, implying an emitting region significantly
out of equipartition (see Figure \ref{fig:Contours}).

An additional point is that the value of $\gamma_{\rm min}$,
constrained between $2.5\times10^4$ and $4.5\times10^4$, is unusually
high compared to standard SSC modeling of blazars.  The fact that the
modeling of hard-VHE-spectrum HSPs requires such a high value of
$\gamma_{\rm min}$ has been previously noted by
\cite{Katarzynski:2006ve}, who claimed that $\gamma_{\rm min} \geq
10^4$ can be a characteristic of this kind of source.
\citeauthor{Katarzynski:2006ve} proposed two alternatives to explain a
high $\gamma_{\rm min}$: either the injected particle population is
characterized by a low-energy cut off and no cooling mechanism is
efficient enough to populate the low-energy part of the spectrum, or
there is an equilibrium between the cooling and the reheating of
particles due to stochastic particle acceleration.

Finally, as stated previously, if the same population of electrons is
responsible for both the X-ray and VHE emission then variability at
X-ray energies should imply VHE variability.  In the SSC scenario a
correlation between X-rays and VHE photons is naturally expected if
the scattering occurs in the Thomson regime. The onset of
Klein-Nishina effects can be computed as a function of the Doppler
factor of the emitting region \citep[see][]{Tavecchio:1998fk}. For the
modeling presented here, assuming $\delta=53$, we obtain that 100 GeV
photons are produced by soft X-rays in the Thomson regime, while hard
X-rays (above 5 keV) are already Comptonized in the Klein-Nishina
regime. On the other hand, for 10 TeV photons the scattering of X-rays
is entirely (above 0.05 keV) happening in the Klein-Nishina
regime. Therefore, we do expect a correlation between X-rays and
$\gamma$-rays if the variability in X-rays is characterized by a
simple variation in the overall normalization of the synchrotron
component; on the other hand, if the peak of the synchrotron component
shifts towards higher energies, this would not affect the measured VHE
flux, nor the spectral index.

\begin{table}
\begin{center}
  \caption{Best-fit SSC model parameters for 1ES\,0229+200.  The
    minimum and maximum values are reported for each parameter.  Note
    that the model parameters are correlated.\label{tab:SED}}
\begin{tabular}{ccccc}
  \tableline\tableline
  Parameter & Units & Current & Kaufmann$^1$ & Tavecchio$^2$ \\
  \tableline
  $\delta$           &                & $53 - 100$   & 40    & 50    \\ 
  $B$                & [$10^{-3}$ G]  & $0.8 - 3.3$ & 0.032 & 0.4   \\
  $R$                & [$10^{15}$ cm] & $4.7 - 29$   & 1000  & 54    \\
  $\gamma_{\rm min}$ & [$10^{4}$]     & $2.5 - 4.5$  & 39    & 50    \\
  $\gamma_{\rm max}$ & [$10^{6}$]     & $3.0 - 7.0$  & 190   & 40    \\
  $\gamma_{\rm b}$   & [$10^{6}$]     & -            & 62    & -     \\
  $K$                & [$10^{-12}$ cm$^{-3}$] & $2.9-180$    & N/A   & N/A \\
  $\alpha$           &                & $2.18 - 2.30$  & 2.6   & 2.85  \\
  \tableline
\end{tabular}
\end{center}
$^1$ \citet{Kaufmann:2011fk}\\
$^2$ \citet{Tavecchio:2009ys}, model 3
\end{table}

\section{Summary and Conclusions}

VERITAS performed a long-term observation of the VHE HSP 1ES\,0229+200
from 2010 to 2012 for a total time of 54.3 hours, providing the most
detailed VHE SED of this blazar to date.  The overall average integral
flux during this time was $(23 \pm 3_{\rm stat} \pm 6_{\rm sys})
\times 10^{-9}$ m$^{-2}$ s$^{-1}$ ($E > $ 300 GeV) and the spectrum is
well described by a power law with photon index $\Gamma = 2.59 \pm
0.12_{\rm stat} \pm 0.26_{\rm sys}$.  The detected VHE emission shows
evidence for variability on yearly time scales (probability of the
flux being constant is 1.6\%), and a period of higher flux was
detected in 2009 October where the integral flux was measured to be
$(42 \pm 6_{\rm stat} \pm 11_{\rm sys}) \times 10^{-9}$ m$^{-2}$
s$^{-1}$ ($E >$ 300 GeV).  No significant change in spectral shape is
seen.

This is the first indication of variability at VHE for this blazar
and, combined with the demonstrated variability of many TeV blazars
and the measured variability at X-ray energies, implies that studies
of the IGMF that depend on a constant flux should not be performed
using this object.  At the very least, the studies must include the
systematic uncertainties inherent in time-averaged SED modeling of
variable sources like 1ES\,1218+304, as suggested by
\citet{Arlen:2012qf}.  The likely detection of variability weakens the
IGMF lower limits based on 1ES\,0229+200 and severely complicates any
IGMF interpretation.

It has been suggested that the photons detected from the direction of
distant ($z >\sim 0.15$) hard-spectrum VHE blazars are actually
secondary $\gamma$-rays produced by the interaction of primary cosmic
rays of energies $10^{16} - 10^{19}$ eV with EBL background photons
\citep{Essey:2010ra,Murase:2012fk}.  This proposal has been used to
explain the detection of distant VHE blazars and to provide a possible
origin of ultra-high-energy cosmic rays \citep{Essey:2012kx}.  Finding
evidence for VHE variability in 1ES\,0229+200 challenges these models
\citep{Prosekin:2012uq} because the reprocessed emission is not
expected to show temporal variability.

The VHE observations were supported by several multi-wavelength data
sets ranging over many orders of magnitude in energy from optical to
GeV.  This allowed for detailed SED modeling based on the code of
\citet{Katarzynski:2001vn}. The best-fit model indicates that the
emission region is relatively small and that the magnetic field is
relatively large compared to previous modeling attempts. The Doppler
factor of $\delta \geq$ 53 is similar to that found by
\citet{Tavecchio:2009ys} and \cite{Weidinger:2010uq}, but is greater
than what was assumed by \citet{Kaufmann:2011fk} (although larger
values could also have been used in that effort).

Since we can now constrain both the synchrotron peak and the
high-energy peak due to the additional BAT and VHE data, we found that
the synchrotron peak is located at a lower frequency than previously
thought.  This means that 1ES\,0229+200 has a lower IC-to-synchrotron
ratio, more in line with the rest of the VHE blazar population.  The
high-energy peak location is similar to that of the VHE HSP
1ES\,1101-232 \citep[$\sim 10^{27}$ Hz,][]{Aharonian:2006oc} but an
order of magnitude higher than that of 1ES\,1218+304 \citep[$3.9
  \times 10^{25}$ Hz,][]{Ruger:2010uq}. These measurements should be
taken as order of magnitude estimates, since the SED coverage of these
two blazars is sparse and both are known to be variable.

The observations of 1ES\,0229+200 presented here are part of the
VERITAS long-term blazar observing program.  This program was
developed to build up a database of SEDs from a variety of blazars.
Under these auspices, we have produced the most detailed SED
measurement of this hard-spectrum distant blazar to date, and we have
discovered evidence for variability at VHE.  Regular VERITAS
observations of 1ES\,0229+200 are continuing which will be used to
further characterize the SED and the nature of the underlying
variability.

\acknowledgements

VERITAS is supported by grants from the U.S. Department of Energy
Office of Science, the U.S. National Science Foundation and the
Smithsonian Institution, by NSERC in Canada, by Science Foundation
Ireland (SFI 10/RFP/AST2748) and by STFC in the U.K. We acknowledge
the excellent work of the technical support staff at the Fred Lawrence
Whipple Observatory and at the collaborating institutions in the
construction and operation of the instrument. This research has made
use of the NASA/IPAC Extragalactic Database (NED) which is operated by
the Jet Propulsion Laboratory, California Institute of Technology,
under contract with the National Aeronautics and Space Administration.


\begin{thebibliography}{}
\expandafter\ifx\csname natexlab\endcsname\relax\def\natexlab#1{#1}\fi

\bibitem[{{Abdo} {et~al.}(2011{\natexlab{a}}){Abdo}, {Ackermann}, {Ajello},
  {Baldini}, {Ballet}, {Barbiellini}, {Bastieri}, {Bechtol}, {Bellazzini},
  {Berenji}, \& et~al.}]{Abdo:2011uq}
{Abdo}, A.~A., {Ackermann}, M., {Ajello}, M., {et~al.} 2011{\natexlab{a}},
  \apj, 736, 131

\bibitem[{{Abdo} {et~al.}(2011{\natexlab{b}}){Abdo}, {Ackermann}, {Ajello},
  {Allafort}, {Baldini}, {Ballet}, {Barbiellini}, {Baring}, {Bastieri},
  {Bechtol}, \& et~al.}]{Abdo:2011tw}
---. 2011{\natexlab{b}}, \apj, 727, 129

\bibitem[{{Abramowski} {et~al.}(2012){Abramowski}, {Acero}, {Aharonian},
  {Akhperjanian}, {Anton}, {Balzer}, {Barnacka}, {Barres de Almeida},
  {Becherini}, \& et~al.}]{Abramowski:2012vn}
{Abramowski}, A., {Acero}, F., {Aharonian}, F., {et~al.} 2012, \aap, 539, A149

\bibitem[{{Abramowski} {et~al.}(2013){Abramowski}, {Acero}, {Aharonian},
  {Akhperjanian}, {Anton}, {Balenderan}, {Balzer}, {Barnacka}, {Becherini},
  {Becker Tjus}, {Bernl{\"o}hr}, {Birsin}, {Biteau}, {Bochow}, {Boisson},
  {Bolmont}, {Bordas}, {Brucker}, {Brun}, {Brun}, {Bulik}, {Carrigan},
  {Casanova}, {Cerruti}, {Chadwick}, {Charbonnier}, {Chaves}, {Cheesebrough},
  {Cologna}, {Conrad}, {Couturier}, {Dalton}, {Daniel}, {Davids}, {Degrange},
  {Deil}, {deWilt}, {Dickinson}, {Djannati-Ata{\"\i}}, {Domainko},
  {O'C.~Drury}, {Dubus}, {Dutson}, {Dyks}, {Dyrda}, {Egberts}, {Eger},
  {Espigat}, {Fallon}, {Farnier}, {Fegan}, {Feinstein}, {Fernandes},
  {Fernandez}, {Fiasson}, {Fontaine}, {F{\"o}rster}, {F{\"u}{\ss}ling},
  {Gajdus}, {Gallant}, {Garrigoux}, {Gast}, {Giebels}, {Glicenstein},
  {Gl{\"u}ck}, {G{\"o}ring}, {Grondin}, {H{\"a}ffner}, {Hague}, {Hahn},
  {Hampf}, {Harris}, {Heinz}, {Heinzelmann}, {Henri}, {Hermann}, {Hillert},
  {Hinton}, {Hofmann}, {Hofverberg}, {Holler}, {Horns}, {Jacholkowska}, {Jahn},
  {Jamrozy}, {Jung}, {Kastendieck}, {Katarzy{\'n}ski}, {Katz}, {Kaufmann},
  {Kh{\'e}lifi}, {Klochkov}, {Klu{\'z}niak}, {Kneiske}, {Komin}, {Kosack},
  {Kossakowski}, {Krayzel}, {Laffon}, {Lamanna}, {Lenain}, {Lennarz}, {Lohse},
  {Lopatin}, {Lu}, {Marandon}, {Marcowith}, {Masbou}, {Maurin}, {Maxted},
  {Mayer}, {McComb}, {Medina}, {M{\'e}hault}, {Menzler}, {Moderski}, {Mohamed},
  {Moulin}, {Naumann}, {Naumann-Godo}, {de Naurois}, {Nedbal}, {Nguyen},
  {Niemiec}, {Nolan}, {Ohm}, {de O{\~n}a Wilhelmi}, {Opitz}, {Ostrowski},
  {Oya}, {Panter}, {Parsons}, {Paz Arribas}, {Pekeur}, {Pelletier}, {Perez},
  {Petrucci}, {Peyaud}, {Pita}, {P{\"u}hlhofer}, {Punch}, {Quirrenbach},
  {Raue}, {Reimer}, {Reimer}, {Renaud}, {de los Reyes}, {Rieger}, {Ripken},
  {Rob}, {Rosier-Lees}, {Rowell}, {Rudak}, {Rulten}, {Sahakian}, {Sanchez},
  {Santangelo}, {Schlickeiser}, {Schulz}, {Schwanke}, {Schwarzburg},
  {Schwemmer}, {Sheidaei}, {Skilton}, {Sol}, {Spengler}, {Stawarz},
  {Steenkamp}, {Stegmann}, {Stinzing}, {Stycz}, {Sushch}, {Szostek},
  {Tavernet}, {Terrier}, {Tluczykont}, {Valerius}, {van Eldik}, {Vasileiadis},
  {Venter}, {Viana}, {Vincent}, {V{\"o}lk}, {Volpe}, {Vorobiov}, {Vorster},
  {Wagner}, {Ward}, {White}, {Wierzcholska}, {Wouters}, {Zacharias}, {Zajczyk},
  {Zdziarski}, {Zech}, \& {Zechlin}}]{Abramowski:2013qf}
---. 2013, \aap, 550, A4

\bibitem[{{Acciari} {et~al.}(2008){Acciari}, {Beilicke}, {Blaylock},
  {Bradbury}, {Buckley}, {Bugaev}, {Butt}, {Byrum}, {Celik}, {Cesarini},
  {Ciupik}, {Chow}, {Cogan}, {Colin}, {Cui}, {Daniel}, {Duke}, {Ergin},
  {Falcone}, {Fegan}, {Finley}, {Fortin}, {Fortson}, {Gall}, {Gibbs},
  {Gillanders}, {Grube}, {Guenette}, {Hanna}, {Hays}, {Holder}, {Horan},
  {Hughes}, {Hui}, {Humensky}, {Kaaret}, {Kieda}, {Kildea}, {Konopelko},
  {Krawczynski}, {Krennrich}, {Lang}, {LeBohec}, {Lee}, {Maier}, {McCann},
  {McCutcheon}, {Millis}, {Moriarty}, {Mukherjee}, {Nagai}, {Ong}, {Pandel},
  {Perkins}, {Pizlo}, {Pohl}, {Quinn}, {Ragan}, {Reynolds}, {Rose},
  {Schroedter}, {Sembroski}, {Smith}, {Steele}, {Swordy}, {Toner}, {Valcarcel},
  {Vassiliev}, {Wagner}, {Wakely}, {Ward}, {Weekes}, {Weinstein}, {White},
  {Williams}, {Wissel}, {Wood}, \& {Zitzer}}]{Acciari:2008oq}
{Acciari}, V.~A., {Beilicke}, M., {Blaylock}, G., {et~al.} 2008, \apj, 679,
  1427

\bibitem[{{Ackermann} {et~al.}(2011){Ackermann},{Ajello}, {Allafort},
    {et~al.}}]{Ackermann:2011fk}
{Ackermann}, M., {Ajello}, M., {Allafort}, A., {et~al.} 2011, \apj,
743, 171

\bibitem[{{Ackermann} {et~al.}(2012){Ackermann}, {Ajello}, {Allafort},
  {Schady}, {Baldini}, {Ballet}, {Barbiellini}, {Bastieri}, {Bellazzini},
  {Blandford}, {Bloom}, {Borgland}, {Bottacini}, {Bouvier}, {Bregeon},
  {Brigida}, {Bruel}, {Buehler}, {Buson}, {Caliandro}, {Cameron}, {Caraveo},
  {Cavazzuti}, {Cecchi}, {Charles}, {Chaves}, {Chekhtman}, {Cheung}, {Chiang},
  {Chiaro}, {Ciprini}, {Claus}, {Cohen-Tanugi}, {Conrad}, {Cutini},
  {D'Ammando}, {de Palma}, {Dermer}, {Digel}, {do Couto e Silva},
  {Dom{\'{\i}}nguez}, {Drell}, {Drlica-Wagner}, {Favuzzi}, {Fegan}, {Focke},
  {Franckowiak}, {Fukazawa}, {Funk}, {Fusco}, {Gargano}, {Gasparrini},
  {Gehrels}, {Germani}, {Giglietto}, {Giordano}, {Giroletti}, {Glanzman},
  {Godfrey}, {Grenier}, {Grove}, {Guiriec}, {Gustafsson}, {Hadasch},
  {Hayashida}, {Hays}, {Jackson}, {Jogler}, {Kataoka}, {Kn{\"o}dlseder},
  {Kuss}, {Lande}, {Larsson}, {Latronico}, {Longo}, {Loparco}, {Lovellette},
  {Lubrano}, {Mazziotta}, {McEnery}, {Mehault}, {Michelson}, {Mizuno}, {Monte},
  {Monzani}, {Morselli}, {Moskalenko}, {Murgia}, {Tramacere}, {Nuss},
  {Greiner}, {Ohno}, {Ohsugi}, {Omodei}, {Orienti}, {Orlando}, {Ormes},
  {Paneque}, {Perkins}, {Pesce-Rollins}, {Piron}, {Pivato}, {Porter},
  {Rain{\`o}}, {Rando}, {Razzano}, {Razzaque}, {Reimer}, {Reimer}, {Reyes},
  {Ritz}, {Rau}, {Romoli}, {Roth}, {S{\'a}nchez-Conde}, {Sanchez}, {Scargle},
  {Sgr{\`o}}, {Siskind}, {Spandre}, {Spinelli}, {Stawarz}, {Suson},
  {Takahashi}, {Tanaka}, {Thayer}, {Thompson}, {Tibaldo}, {Tinivella},
  {Torres}, {Tosti}, {Troja}, {Usher}, {Vandenbroucke}, {Vasileiou},
  {Vianello}, {Vitale}, {Waite}, {Winer}, {Wood}, \& {Wood}}]{Ackermann:2012ly}
{Ackermann}, M., {Ajello}, M., {Allafort}, A., {et~al.} 2012, Science, 338,
  1190

\bibitem[{{Aharonian} {et~al.}(2003){Aharonian}, {Akhperjanian},
    {Beilicke}, {et.~al}}]{Aharonian:2003bs}
{Aharonian}, F.~A., {Akhperjanian}, A., {Beilicke}, M. , {et~al.} 2003
\aap, 403, 523

\bibitem[{{Aharonian} {et~al.}(2004){Aharonian}, {Akhperjanian}, {Beilicke},
  {Bernl{\"o}hr}, {B{\"o}rst}, {Bojahr}, {Bolz}, {Coarasa}, {Contreras},
  {Cortina}, {Denninghoff}, {Fonseca}, {Girma}, {G{\"o}tting}, {Heinzelmann},
  {Hermann}, {Heusler}, {Hofmann}, {Horns}, {Jung}, {Kankanyan}, {Kestel},
  {Konopelko}, {Kornmeyer}, {Kranich}, {Lampeitl}, {Lopez}, {Lorenz},
  {Lucarelli}, {Mang}, {Mazin}, {Meyer}, {Mirzoyan}, {Moralejo},
  {Ona-Wilhelmi}, {Panter}, {Plyasheshnikov}, {P{\"u}hlhofer}, {de los Reyes},
  {Rhode}, {Ripken}, {Rowell}, {Sahakian}, {Samorski}, {Schilling}, {Siems},
  {Sobzynska}, {Stamm}, {Tluczykont}, {Vitale}, {V{\"o}lk}, {Wiedner}, \&
  {Wittek}}]{Aharonian:2004pi}
{Aharonian}, F., {Akhperjanian}, A., {Beilicke}, M., {et~al.} 2004, \aap, 421,
  529

\bibitem[{{Aharonian} {et~al.}(2006){Aharonian}, {Akhperjanian}, {Bazer-Bachi},
  {Beilicke}, {Benbow}, {Berge}, {Bernl{\"o}hr}, {Boisson}, {Bolz}, {Borrel},
  {Braun}, {Breitling}, {Brown}, {B{\"u}hler}, {B{\"u}sching}, {Carrigan},
  {Chadwick}, {Chounet}, {Cornils}, {Costamante}, {Degrange}, {Dickinson},
  {Djannati-Ata{\"\i}}, {O'C.~Drury}, {Dubus}, {Egberts}, {Emmanoulopoulos},
  {Espigat}, {Feinstein}, {Ferrero}, {Fiasson}, {Fontaine}, {Funk}, {Funk},
  {Gallant}, {Giebels}, {Glicenstein}, {Goret}, {Hadjichristidis}, {Hauser},
  {Hauser}, {Heinzelmann}, {Henri}, {Hermann}, {Hinton}, {Hofmann}, {Holleran},
  {Horns}, {Jacholkowska}, {de Jager}, {Kh{\'e}lifi}, {Komin}, {Konopelko},
  {Kosack}, {Latham}, {Le Gallou}, {Lemi{\`e}re}, {Lemoine-Goumard}, {Lohse},
  {Martin}, {Martineau-Huynh}, {Marcowith}, {Masterson}, {McComb}, {de
  Naurois}, {Nedbal}, {Nolan}, {Noutsos}, {Orford}, {Osborne}, {Ouchrif},
  {Panter}, {Pelletier}, {Pita}, {P{\"u}hlhofer}, {Punch}, {Raubenheimer},
  {Raue}, {Rayner}, {Reimer}, {Reimer}, {Ripken}, {Rob}, {Rolland}, {Rowell},
  {Sahakian}, {Saug{\'e}}, {Schlenker}, {Schlickeiser}, {Schwanke}, {Sol},
  {Spangler}, {Spanier}, {Steenkamp}, {Stegmann}, {Superina}, {Tavernet},
  {Terrier}, {Th{\'e}oret}, {Tluczykont}, {van Eldik}, {Vasileiadis}, {Venter},
  {Vincent}, {V{\"o}lk}, {Wagner}, \& {Ward}}]{Aharonian:2006id}
{Aharonian}, F., {Akhperjanian}, A.~G., {Bazer-Bachi}, A.~R., {et~al.} 2006,
  Astronomy and Astrophysics, 457, 899

\bibitem[{{Aharonian} {et~al.}(2007{\natexlab{a}}){Aharonian}, {Akhperjanian},
  {Bazer-Bachi}, {Beilicke}, {Benbow}, {Berge}, {Bernl{\"o}hr}, {Boisson},
  {Bolz}, {Borrel}, {Braun}, {Brion}, {Brown}, {B{\"u}hler}, {B{\"u}sching},
  {Boutelier}, {Carrigan}, {Chadwick}, {Chounet}, {Coignet}, {Cornils},
  {Costamante}, {Degrange}, {Dickinson}, {Djannati-Ata{\"\i}}, {O'C.~Drury},
  {Dubus}, {Egberts}, {Emmanoulopoulos}, {Espigat}, {Farnier}, {Feinstein},
  {Ferrero}, {Fiasson}, {Fontaine}, {Funk}, {Funk}, {F{\"u}{\ss}ling},
  {Gallant}, {Giebels}, {Glicenstein}, {Gl{\"u}ck}, {Goret}, {Hadjichristidis},
  {Hauser}, {Hauser}, {Heinzelmann}, {Henri}, {Hermann}, {Hinton}, {Hoffmann},
  {Hofmann}, {Holleran}, {Hoppe}, {Horns}, {Jacholkowska}, {de Jager},
  {Kendziorra}, {Kerschhaggl}, {Kh{\'e}lifi}, {Komin}, {Kosack}, {Lamanna},
  {Latham}, {Le Gallou}, {Lemi{\`e}re}, {Lemoine-Goumard}, {Lohse}, {Martin},
  {Martineau-Huynh}, {Marcowith}, {Masterson}, {Maurin}, {McComb}, {Moulin},
  {de Naurois}, {Nedbal}, {Nolan}, {Noutsos}, {Olive}, {Orford}, {Osborne},
  {Panter}, {Pelletier}, {Petrucci}, {Pita}, {P{\"u}hlhofer}, {Punch},
  {Ranchon}, {Raubenheimer}, {Raue}, {Rayner}, {Ripken}, {Rob}, {Rolland},
  {Rosier-Lees}, {Rowell}, {Sahakian}, {Santangelo}, {Saug{\'e}}, {Schlenker},
  {Schlickeiser}, {Schr{\"o}der}, {Schwanke}, {Schwarzburg}, {Schwemmer},
  {Shalchi}, {Sol}, {Spangler}, {Spanier}, {Steenkamp}, {Stegmann}, {Superina},
  {Tam}, {Tavernet}, {Terrier}, {Tluczykont}, {van Eldik}, {Vasileiadis},
  {Venter}, {Vialle}, {Vincent}, {V{\"o}lk}, {Wagner}, \&
  {Ward}}]{Aharonian:2007kx}
---. 2007{\natexlab{a}}, \aap, 470, 475

\bibitem[{{Aharonian} {et~al.}(2007{\natexlab{b}}){Aharonian}, {Akhperjanian},
  {Barres de Almeida}, {Bazer-Bachi}, {Behera}, {Beilicke}, {Benbow},
  {Bernl{\"o}hr}, {Boisson}, {Bolz}, {Borrel}, {Braun}, {Brion}, {Brown},
  {B{\"u}hler}, {Bulik}, {B{\"u}sching}, {Boutelier}, {Carrigan}, {Chadwick},
  {Chounet}, {Clapson}, {Coignet}, {Cornils}, {Costamante}, {Dalton},
  {Degrange}, {Dickinson}, {Djannati-Ata{\"\i}}, {Domainko}, {O'C.~Drury},
  {Dubois}, {Dubus}, {Dyks}, {Egberts}, {Emmanoulopoulos}, {Espigat},
  {Farnier}, {Feinstein}, {Fiasson}, {F{\"o}rster}, {Fontaine}, {Funk},
  {F{\"u}{\ss}ling}, {Gallant}, {Giebels}, {Glicenstein}, {Gl{\"u}ck}, {Goret},
  {Hadjichristidis}, {Hauser}, {Hauser}, {Heinzelmann}, {Henri}, {Hermann},
  {Hinton}, {Hoffmann}, {Hofmann}, {Holleran}, {Hoppe}, {Horns},
  {Jacholkowska}, {de Jager}, {Jung}, {Katarzy{\'n}ski}, {Kendziorra},
  {Kerschhaggl}, {Kh{\'e}lifi}, {Keogh}, {Komin}, {Kosack}, {Lamanna},
  {Latham}, {Lemi{\`e}re}, {Lemoine-Goumard}, {Lenain}, {Lohse}, {Martin},
  {Martineau-Huynh}, {Marcowith}, {Masterson}, {Maurin}, {Maurin}, {McComb},
  {Moderski}, {Moulin}, {de Naurois}, {Nedbal}, {Nolan}, {Ohm}, {Olive}, {de
  O{\~n}a Wilhelmi}, {Orford}, {Osborne}, {Ostrowski}, {Panter}, {Pedaletti},
  {Pelletier}, {Petrucci}, {Pita}, {P{\"u}hlhofer}, {Punch}, {Ranchon},
  {Raubenheimer}, {Raue}, {Rayner}, {Renaud}, {Ripken}, {Rob}, {Rolland},
  {Rosier-Lees}, {Rowell}, {Rudak}, {Ruppel}, {Sahakian}, {Santangelo},
  {Schlickeiser}, {Sch{\"o}ck}, {Schr{\"o}der}, {Schwanke}, {Schwarzburg},
  {Schwemmer}, {Shalchi}, {Sol}, {Spangler}, {Stawarz}, {Steenkamp},
  {Stegmann}, {Superina}, {Tam}, {Tavernet}, {Terrier}, {van Eldik},
  {Vasileiadis}, {Venter}, {Vialle}, {Vincent}, {Vivier}, {V{\"o}lk}, {Volpe},
  {Wagner}, {Ward}, {Zdziarski}, \& {Zech}}]{Aharonian:2007vn}
{Aharonian}, F., {Akhperjanian}, A.~G., {Barres de Almeida}, U., {et~al.}
  2007{\natexlab{b}}, \aap, 473, L25

\bibitem[{{Aharonian} {et~al.}(2007{\natexlab{c}}){Aharonian}, {Akhperjanian},
  {Barres de Almeida}, {Bazer-Bachi}, {Behera}, {Beilicke}, {Benbow},
  {Bernl{\"o}hr}, {Boisson}, {Bolz}, {Borrel}, {Braun}, {Brion}, {Brown},
  {B{\"u}hler}, {Bulik}, {B{\"u}sching}, {Boutelier}, {Carrigan}, {Chadwick},
  {Chounet}, {Clapson}, {Coignet}, {Cornils}, {Costamante}, {Dalton},
  {Degrange}, {Dickinson}, {Djannati-Ata{\"\i}}, {Domainko}, {O'C.~Drury},
  {Dubois}, {Dubus}, {Dyks}, {Egberts}, {Emmanoulopoulos}, {Espigat},
  {Farnier}, {Feinstein}, {Fiasson}, {F{\"o}rster}, {Fontaine}, {Funk},
  {F{\"u}{\ss}ling}, {Gallant}, {Giebels}, {Glicenstein}, {Gl{\"u}ck}, {Goret},
  {Hadjichristidis}, {Hauser}, {Hauser}, {Heinzelmann}, {Henri}, {Hermann},
  {Hinton}, {Hoffmann}, {Hofmann}, {Holleran}, {Hoppe}, {Horns},
  {Jacholkowska}, {de Jager}, {Jung}, {Katarzy{\'n}ski}, {Kendziorra},
  {Kerschhaggl}, {Kh{\'e}lifi}, {Keogh}, {Komin}, {Kosack}, {Lamanna},
  {Latham}, {Lemi{\`e}re}, {Lemoine-Goumard}, {Lenain}, {Lohse}, {Martin},
  {Martineau-Huynh}, {Marcowith}, {Masterson}, {Maurin}, {Maurin}, {McComb},
  {Moderski}, {Moulin}, {de Naurois}, {Nedbal}, {Nolan}, {Ohm}, {Olive}, {de
  O{\~n}a Wilhelmi}, {Orford}, {Osborne}, {Ostrowski}, {Panter}, {Pedaletti},
  {Pelletier}, {Petrucci}, {Pita}, {P{\"u}hlhofer}, {Punch}, {Ranchon},
  {Raubenheimer}, {Raue}, {Rayner}, {Renaud}, {Ripken}, {Rob}, {Rolland},
  {Rosier-Lees}, {Rowell}, {Rudak}, {Ruppel}, {Sahakian}, {Santangelo},
  {Schlickeiser}, {Sch{\"o}ck}, {Schr{\"o}der}, {Schwanke}, {Schwarzburg},
  {Schwemmer}, {Shalchi}, {Sol}, {Spangler}, {Stawarz}, {Steenkamp},
  {Stegmann}, {Superina}, {Tam}, {Tavernet}, {Terrier}, {van Eldik},
  {Vasileiadis}, {Venter}, {Vialle}, {Vincent}, {Vivier}, {V{\"o}lk}, {Volpe},
  {Wagner}, {Ward}, {Zdziarski}, \& {Zech}}]{Aharonian:2007dn}
---. 2007{\natexlab{c}}, \aap, 475, L9

\bibitem[{Aharonian {et~al.}(2006)Aharonian, Akhperjanian, Bazer-Bachi,
  Beilicke, Benbow, Berge, Bernl\"{o}hr, Boisson, Bolz, Borrel, Braun,
  Breitling, Brown, Chadwick, Chounet, Cornils, Costamante, Degrange,
  Dickinson, Djannati-Ata\"{\i}, Drury, Dubus, Emmanoulopoulos, Espigat,
  Feinstein, Fontaine, Fuchs, Funk, Gallant, Giebels, Gillessen, Glicenstein,
  Goret, Hadjichristidis, Hauser, Hauser, Heinzelmann, Henri, Hermann, Hinton,
  Hofmann, Holleran, Horns, Jacholkowska, de~Jager, Kh\'{e}lifi, Klages, Komin,
  Konopelko, Latham, {Le Gallou}, Lemi\`{e}re, Lemoine-Goumard, Leroy, Lohse,
  Martin, Martineau-Huynh, Marcowith, Masterson, McComb, de~Naurois, Nolan,
  Noutsos, Orford, Osborne, Ouchrif, Panter, Pelletier, Pita, P\"{u}hlhofer,
  Punch, Raubenheimer, Raue, Raux, Rayner, Reimer, Reimer, Ripken, Rob,
  Rolland, Rowell, Sahakian, Saug\'{e}, Schlenker, Schlickeiser, Schuster,
  Schwanke, Siewert, Sol, Spangler, Steenkamp, Stegmann, Tavernet, Terrier,
  Th\'{e}oret, Tluczykont, van Eldik, Vasileiadis, Venter, Vincent, V\"{o}lk,
  \& Wagner}]{Aharonian:2006oc}
Aharonian, F.~A., Akhperjanian, A.~G., Bazer-Bachi, A.~R., {et~al.} 2006,
  Nature, 440, 1018

\bibitem[{{Albert} {et~al.}(2006){Albert}, {Aliu}, {Anderhub}, {Antoranz},
  {Armada}, {Asensio}, {Baixeras}, {Barrio}, {Bartelt}, {Bartko}, {Bastieri},
  {Bavikadi}, {Bednarek}, {Berger}, {Bigongiari}, {Biland}, {Bisesi}, {Bock},
  {Bretz}, {Britvitch}, {Camara}, {Chilingarian}, {Ciprini}, {Coarasa},
  {Commichau}, {Contreras}, {Cortina}, {Curtef}, {Danielyan}, {Dazzi}, {De
  Angelis}, {de los Reyes}, {De Lotto}, {Domingo-Santamar{\'{\i}}a}, {Dorner},
  {Doro}, {Errando}, {Fagiolini}, {Ferenc}, {Fern{\'a}ndez}, {Firpo}, {Flix},
  {Fonseca}, {Font}, {Galante}, {Garczarczyk}, {Gaug}, {Giller}, {Goebel},
  {Hakobyan}, {Hayashida}, {Hengstebeck}, {H{\"o}hne}, {Hose}, {Jacon},
  {Kalekin}, {Kranich}, {Laille}, {Lenisa}, {Liebing}, {Lindfors}, {Longo},
  {L{\'o}pez}, {L{\'o}pez}, {Lorenz}, {Lucarelli}, {Majumdar}, {Maneva},
  {Mannheim}, {Mariotti}, {Mart{\'{\i}}nez}, {Mase}, {Mazin}, {Meucci},
  {Meyer}, {Miranda}, {Mirzoyan}, {Mizobuchi}, {Moralejo}, {Nilsson},
  {O{\~n}a-Wilhelmi}, {Ordu{\~n}a}, {Otte}, {Oya}, {Paneque}, {Paoletti},
  {Pasanen}, {Pascoli}, {Pauss}, {Pavel}, {Pegna}, {Persic}, {Peruzzo},
  {Piccioli}, {Poller}, {Prandini}, {Rhode}, {Rico}, {Riegel}, {Rissi},
  {Robert}, {R{\"u}gamer}, {Saggion}, {S{\'a}nchez}, {Sartori}, {Scalzotto},
  {Schmitt}, {Schweizer}, {Shayduk}, {Shinozaki}, {Shore}, {Sidro},
  {Sillanp{\"a}{\"a}}, {Sobczy{\'n}ska}, {Stamerra}, {Stark}, {Takalo},
  {Temnikov}, {Tescaro}, {Teshima}, {Tonello}, {Torres}, {Torres}, {Turini},
  {Vankov}, {Vardanyan}, {Vitale}, {Wagner}, {Wibig}, {Wittek}, \&
  {Zapatero}}]{Albert:2006uv}
{Albert}, J., {Aliu}, E., {Anderhub}, H., {et~al.} 2006, \apjl, 642, L119

\bibitem[{{Arlen} \& {Vassiliev}(2012)}]{Arlen:2012kx}
{Arlen}, T.~C., \& {Vassiliev}, V.~V. 2012, in American Institute of Physics
  Conference Series, Vol. 1505, American Institute of Physics Conference
  Series, ed. F.~A. {Aharonian}, W.~{Hofmann}, \& F.~M. {Rieger}, 606--609

\bibitem[{{Arlen} {et~al.}(2012){Arlen}, {Vassiliev}, {Weisgarber}, {Wakely},
  \& {Yusef Shafi}}]{Arlen:2012qf}
{Arlen}, T.~C., {Vassiliev}, V.~V., {Weisgarber}, T., {Wakely}, S.~P., \&
  {Yusef Shafi}, S. 2012, ArXiv e-prints, 1210.2802

\bibitem[{{Baumgartner} {et~al.}(2013){Baumgartner}, {Tueller}, {Markwardt},
  {Skinner}, {Barthelmy}, {Mushotzky}, {Evans}, \&
  {Gehrels}}]{Baumgartner:2012ly}
{Baumgartner}, W.~H., {Tueller}, J., {Markwardt}, C.~B., {et~al.}
2013, ApJS, 207, 19

\bibitem[{{B{\"o}ttcher}(2010)}]{Boettcher:2010fk}
{B{\"o}ttcher}, M. 2010, in Fermi Meets Jansky - AGN at Radio and Gamma-Rays,
  ed. T.~{Savolainen}, E.~{Ros}, R.~W. {Porcas}, \& J.~A. {Zensus}

\bibitem[{{Bradt} {et~al.}(1993){Bradt}, {Rothschild}, \&
  {Swank}}]{Bradt:1993fk}
{Bradt}, H.~V., {Rothschild}, R.~E., \& {Swank}, J.~H. 1993, \aaps, 97, 355

\bibitem[{{Broderick} {et~al.}(2012){Broderick}, {Chang}, \&
      {Pfrommer}}]{Broderick:2012uq}
{Broderick}, A.~E., {Chang}, P. \& {Pfrommer}, C., 2012, \apj, 752, 22

\bibitem[{{Burrows} {et~al.}(2005){Burrows}, {Hill}, {Nousek}, {Kennea},
  {Wells}, {Osborne}, {Abbey}, {Beardmore}, {Mukerjee}, {Short}, {Chincarini},
  {Campana}, {Citterio}, {Moretti}, {Pagani}, {Tagliaferri}, {Giommi},
  {Capalbi}, {Tamburelli}, {Angelini}, {Cusumano}, {Br{\"a}uninger}, {Burkert},
  \& {Hartner}}]{Burrows:2005yg}
{Burrows}, D.~N., {Hill}, J.~E., {Nousek}, J.~A., {et~al.} 2005, Space Science
  Reviews, 120, 165

\bibitem[{{Cerruti} {et~al.}(2013){Cerruti}, {Boisson}, \&
  {Zech}}]{Cerruti:2013zr}
{Cerruti}, M., {Boisson}, C., \& {Zech}, A. 2013, \aap, 558, A47

\bibitem[{{Coppi} \& {Aharonian}(1998)}]{Coppi:1998fk}
{Coppi}, P.~S., \& {Aharonian}, F.~A. 1998, in 19th Texas Symposium on
  Relativistic Astrophysics and Cosmology, ed. J.~{Paul}, T.~{Montmerle}, \&
  E.~{Aubourg}

\bibitem[{{Costamante} \& {Ghisellini}(2002)}]{Costamante:2002dq}
{Costamante}, L., \& {Ghisellini}, G. 2002, \aap, 384, 56

\bibitem[{{de la Calle P{\'e}rez} {et~al.}(2003){de la Calle P{\'e}rez},
  {Bond}, {Boyle}, {Bradbury}, {Buckley}, {Carter-Lewis}, {Celik}, {Cui},
  {Dowdall}, {Duke}, {Falcone}, {Fegan}, {Fegan}, {Finley}, {Fortson},
  {Gaidos}, {Gibbs}, {Gammell}, {Hall}, {Hall}, {Hillas}, {Holder}, {Horan},
  {Jordan}, {Kertzman}, {Kieda}, {Kildea}, {Knapp}, {Kosack}, {Krawczynski},
  {Krennrich}, {LeBohec}, {Linton}, {Lloyd-Evans}, {Moriarty}, {M{\"u}ller},
  {Nagai}, {Ong}, {Page}, {Pallassini}, {Petry}, {Power-Mooney}, {Quinn},
  {Rebillot}, {Reynolds}, {Rose}, {Schroedter}, {Sembroski}, {Swordy},
  {Vassiliev}, {Wakely}, {Walker}, \& {Weekes}}]{de-la-Calle-Perez:2003cr}
{de la Calle P{\'e}rez}, I., {Bond}, I.~H., {Boyle}, P.~J., {et~al.} 2003,
  \apj, 599, 909

\bibitem[{{Dermer} {et~al.}(2011){Dermer}, {Cavadini}, {Razzaque}, {Finke},
  {Chiang}, \& {Lott}}]{Dermer:2011fk}
{Dermer}, C.~D., {Cavadini}, M., {Razzaque}, S., {et~al.} 2011, \apjl, 733, L21

\bibitem[{{Dolag} {et~al.}(2011){Dolag}, {Dolag}, {Kachelriess},
    {Ostapchenko}, \& {Tom\`{a}s}}]{Dolag:2011uq}
{Dolag}, K., {Kachelriess}, M., {Ostapchenko}, S. \& {Tom\`{a}s},
R. 2011, \apjl, 727, L4

\bibitem[{{Elbaz} {et~al.}(2002){Elbaz}, {Cesarsky}, {Chanial}, {Aussel},
  {Franceschini}, {Fadda}, \& {Chary}}]{Elbaz:2002zr}
{Elbaz}, D., {Cesarsky}, C.~J., {Chanial}, P., {et~al.} 2002, \aap, 384, 848

\bibitem[{{Elvis} {et~al.}(1992){Elvis}, {Plummer}, {Schachter}, \&
  {Fabbiano}}]{Elvis:1992kx}
{Elvis}, M., {Plummer}, D., {Schachter}, J., \& {Fabbiano}, G. 1992, \apjs, 80,
  257

\bibitem[{{Essey} {et~al.}(2010){Essey}, {Kalashev}, {Kusenko}, \&
  {Beacom}}]{Essey:2010ra}
{Essey}, W., {Kalashev}, O.~E., {Kusenko}, A., \& {Beacom}, J.~F. 2010,
  Physical Review Letters, 104, 141102

\bibitem[{{Essey} \& {Kusenko}(2012)}]{Essey:2012kx}
{Essey}, W., \& {Kusenko}, A. 2012, \apjl, 751, L11

\bibitem[{{Falomo} {et~al.}(2000){Falomo}, {Scarpa}, {Treves}, \&
  {Urry}}]{Falomo:2000rs}
{Falomo}, R., {Scarpa}, R., {Treves}, A., \& {Urry}, C.~M. 2000, \apj, 542, 731

\bibitem[{{Fazio} {et~al.}(2004){Fazio}, {Ashby}, {Barmby}, {Hora}, {Huang},
  {Pahre}, {Wang}, {Willner}, {Arendt}, {Moseley}, {Brodwin}, {Eisenhardt},
  {Stern}, {Tollestrup}, \& {Wright}}]{Fazio:2004ka}
{Fazio}, G.~G., {Ashby}, M.~L.~N., {Barmby}, P., {et~al.} 2004, \apjs, 154, 39

\bibitem[{{Fomin} {et~al.}(1994){Fomin}, {Stepanian}, {Lamb}, {Lewis}, {Punch},
  \& {Weekes}}]{Fomin:1994kl}
{Fomin}, V.~P., {Stepanian}, A.~A., {Lamb}, R.~C., {et~al.} 1994, Astroparticle
  Physics, 2, 137

\bibitem[{{Franceschini} {et~al.}(2008){Franceschini}, {Rodighiero}, \&
  {Vaccari}}]{Franceschini:2008wo}
{Franceschini}, A., {Rodighiero}, G., \& {Vaccari}, M. 2008, \aap, 487, 837

\bibitem[{{Georganopoulos} {et~al.}(2010){Georganopoulos}, {Finke}, \&
  {Reyes}}]{Georganopoulos:2010ys}
{Georganopoulos}, M., {Finke}, J.~D., \& {Reyes}, L.~C. 2010, \apjl, 714, L157

\bibitem[{{Gould} \& {Schr{\'e}der}(1967)}]{Gould:1967uf}
{Gould}, R.~J., \& {Schr{\'e}der}, G.~P. 1967, Physical Review, 155, 1408

\bibitem[{{Henri} \& {Saug{\'e}}(2006)}]{Henri:2006}
{Henri}, G., \& {Saug{\'e}}, L. 2006, \apj, 640, 185

\bibitem[{{Hillas}(1985)}]{Hillas:1985vc}
{Hillas}, A.~M. 1985, in International Cosmic Ray Conference, Vol.~3,
  International Cosmic Ray Conference, ed. F.~C. {Jones}, 445--448

\bibitem[{{Holder} {et~al.}(2008){Holder}, {Acciari}, {Aliu},
    {et.~al}}]{Holder:2008sz} 
{Holder}, J., {Acciari}, V.~A., {Aliu}, E., {et~al.} 2008, in American
Institute of Physics Conference Series, Vol. 1085, American Institute
of Physics Conference Series, ed. F.~A.~{Aharonian}, W.~{Hofmann}, \&
F.~{Rieger}, 657--660

\bibitem[{{Horan} {et~al.}(2002){Horan}, {Badran}, {Bond}, {Bradbury},
  {Buckley}, {Carson}, {Carter-Lewis}, {Catanese}, {Cui}, {Dunlea}, {Das}, {de
  la Calle Perez}, {D'Vali}, {Fegan}, {Fegan}, {Finley}, {Gaidos}, {Gibbs},
  {Gillanders}, {Hall}, {Hillas}, {Holder}, {Jordan}, {Kertzman}, {Kieda},
  {Kildea}, {Knapp}, {Kosack}, {Krennrich}, {Lang}, {LeBohec}, {Lessard},
  {Lloyd-Evans}, {McKernan}, {Moriarty}, {Muller}, {Ong}, {Pallassini},
  {Petry}, {Quinn}, {Reay}, {Reynolds}, {Rose}, {Sembroski}, {Sidwell},
  {Stanton}, {Swordy}, {Vassiliev}, {Wakely}, \& {Weekes}}]{Horan:2002fk}
{Horan}, D., {Badran}, H.~M., {Bond}, I.~H., {et~al.} 2002, \apj, 571, 753

\bibitem[{{Huan} {et~al.}(2011){Huan}, {Weisgarber}, {Arlen}, \&
  {Wakely}}]{Huan:2011vn}
{Huan}, H., {Weisgarber}, T., {Arlen}, T., \& {Wakely}, S.~P. 2011, \apjl, 735,
  L28

\bibitem[{{Katarzy{\'n}ski} {et~al.}(2006){Katarzy{\'n}ski}, {Ghisellini},
  {Tavecchio}, {Gracia}, \& {Maraschi}}]{Katarzynski:2006ve}
{Katarzy{\'n}ski}, K., {Ghisellini}, G., {Tavecchio}, F., {Gracia}, J., \&
  {Maraschi}, L. 2006, \mnras, 368, L52

\bibitem[{{Katarzy{\'n}ski} {et~al.}(2001){Katarzy{\'n}ski}, {Sol}, \&
  {Kus}}]{Katarzynski:2001vn}
{Katarzy{\'n}ski}, K., {Sol}, H., \& {Kus}, A. 2001, \aap, 367, 809

\bibitem[{{Kaufmann} {et~al.}(2011){Kaufmann}, {Wagner}, {Tibolla}, \&
  {Hauser}}]{Kaufmann:2011fk}
{Kaufmann}, S., {Wagner}, S.~J., {Tibolla}, O., \& {Hauser}, M. 2011, \aap,
  534, A130

\bibitem[{Kneiske {et~al.}(2002)Kneiske, Mannheim, \&
  Hartmann}]{Kneiske:2002ib}
Kneiske, T.~M., Mannheim, K., \& Hartmann, D.~H. 2002, \aap, 386, 1

\bibitem[{{Lister} {et~al.}(2013){Lister}, {Aller}, {Aller}, {Homan},
      {Kellermann}, {Kovalev}, {Pushkarev}, {Richards}, {Ros}, \&
      {Savolainen}}]{Lister:2013}
{Lister}, M.~L., {Aller}, M.~F., {Aller}, H.~D., {et~al.} 2013, \aj,
146, 120

\bibitem[{{Miniati} \& {Elyiv}(2013)}]{Miniati:2013zr}
{Miniati}, F. and {Elyiv}, A. 2013, \apj, 770, 54

\bibitem[{{Murase} {et~al.}(2012){Murase}, {Dermer}, {Takami}, \&
  {Migliori}}]{Murase:2012fk}
{Murase}, K., {Dermer}, C.~D., {Takami}, H., \& {Migliori}, G. 2012, \apj, 749,
  63

\bibitem[{{Neronov} \& {Semikoz}(2009)}]{Neronov:2009fk}
{Neronov}, A., \& {Semikoz}, D.~V. 2009, \prd, 80, 123012

\bibitem[{{Neronov} \& {Vovk}(2010)}]{Neronov:2010uq}
{Neronov}, A., \& {Vovk}, I. 2010, Science, 328, 73

\bibitem[{{Orr} {et~al.}(2011){Orr}, {Krennrich}, \& {Dwek}}]{Orr:2011uq}
{Orr}, M.~R., {Krennrich}, F., \& {Dwek}, E. 2011, \apj, 733, 77

\bibitem[{{Poole} {et~al.}(2008){Poole}, {Breeveld}, {Page}, {Landsman},
  {Holland}, {Roming}, {Kuin}, {Brown}, {Gronwall}, {Hunsberger}, {Koch},
  {Mason}, {Schady}, {vanden Berk}, {Blustin}, {Boyd}, {Broos}, {Carter},
  {Chester}, {Cucchiara}, {Hancock}, {Huckle}, {Immler}, {Ivanushkina},
  {Kennedy}, {Marshall}, {Morgan}, {Pandey}, {de Pasquale}, {Smith}, \&
  {Still}}]{Poole:2008do}
{Poole}, T.~S., {Breeveld}, A.~A., {Page}, M.~J., {et~al.} 2008, \mnras, 383,
  627

\bibitem[{{Prosekin} {et~al.}(2012){Prosekin}, {Essey}, {Kusenko}, \&
  {Aharonian}}]{Prosekin:2012uq}
{Prosekin}, A., {Essey}, W., {Kusenko}, A., \& {Aharonian}, F. 2012, \apj, 757,
  183

\bibitem[{{Raue} \& {Mazin}(2008)}]{Raue:2008fk}
{Raue}, M., \& {Mazin}, D. 2008, International Journal of Modern Physics D, 17,
  1515

\bibitem[{{Rolke} \& {L{\'o}pez}(2001)}]{Rolke:2001fk}
{Rolke}, W.~A., \& {L{\'o}pez}, A.~M. 2001, Nuclear Instruments and Methods in
  Physics Research A, 458, 745

\bibitem[{{R{\"u}ger} {et~al.}(2010){R{\"u}ger}, {Spanier}, \&
  {Mannheim}}]{Ruger:2010uq}
{R{\"u}ger}, M., {Spanier}, F., \& {Mannheim}, K. 2010, \mnras, 401, 973

\bibitem[{{Schachter} {et~al.}(1993){Schachter}, {Stocke}, {Perlman}, {Elvis},
  {Remillard}, {Granados}, {Luu}, {Huchra}, {Humphreys}, {Urry}, \&
  {Wallin}}]{Schachter:1993uq}
{Schachter}, J.~F., {Stocke}, J.~T., {Perlman}, E., {et~al.} 1993, \apj, 412,
  541

\bibitem[{{Schlegel} {et~al.}(1998){Schlegel}, {Finkbeiner}, \&
  {Davis}}]{Schlegel:1998lw}
{Schlegel}, D.~J., {Finkbeiner}, D.~P., \& {Davis}, M. 1998, \apj, 500, 525

\bibitem[{{Schlickeiser} {et~al.}(2012){Schlickeiser}, {Elyiv},
    {Ibscher} \& {Miniati}}]{Schlickeiser:2012ys} 
{Schlickeiser}, R., {Elyiv}, A., {Ibscher}, D. \& {Miniati}, F. 2012,
\apj, 758, 101

\bibitem[{{Seaton}(1979)}]{Seaton:1979kx}
{Seaton}, M.~J. 1979, \mnras, 187, 73P

\bibitem[{{Stecker} {et~al.}(1992){Stecker}, {de Jager}, \&
  {Salamon}}]{Stecker:1992fk}
{Stecker}, F.~W., {de Jager}, O.~C., \& {Salamon}, M.~H. 1992, \apjl, 390, L49

\bibitem[{{Stecker} {et~al.}(1996){Stecker}, {de Jager}, \&
  {Salamon}}]{Stecker:1996bh}
---. 1996, \apjl, 473, L75

\bibitem[{{Stecker} {et~al.}(2006){Stecker}, {Malkan}, \&
  {Scully}}]{Stecker:2006kx}
{Stecker}, F.~W., {Malkan}, M.~A., \& {Scully}, S.~T. 2006, \apj, 648, 774

\bibitem[{{Taylor}}(2011){Taylor}, {Taylor}]{Taylor:2011fk}
{Taylor}, A.~M. 2011, in Cosmic Rays for Particle and Astroparticle
Physics, ed. S.~{Giani}, C.~{Leroy} \& P.~G.~{Rancoita}, 563--568

\bibitem[{{Tavecchio} {et~al.}(1998){Tavecchio}, {Maraschi}, \&
    {Ghisellini}}]{Tavecchio:1998fk}
{Tavecchio}, F., {Maraschi}, L., \& {Ghisellini}, G. 1998, \apj, 509,
608

\bibitem[{{Tavecchio} {et~al.}(2009){Tavecchio}, {Ghisellini}, {Ghirlanda},
  {Costamante}, \& {Franceschini}}]{Tavecchio:2009ys}
{Tavecchio}, F., {Ghisellini}, G., {Ghirlanda}, G., {Costamante}, L., \&
  {Franceschini}, A. 2009, \mnras, 399, L59

\bibitem[{{Vaughan} {et~al.}(2003){Vaughan}, {Edelson}, {Warwick},\&
    {Uttley}}]{Vaughan:2003fk} 
{Vaughan}, S., {Edelson}, R., {Warwick}, R.~S, \& {Uttley}, P. 2003,
  \mnras, 345, 1271

\bibitem[{{Vovk} {et~al.}(2012){Vovk}, {Taylor}, {Semikoz}, \&
  {Neronov}}]{Vovk:2012vn}
{Vovk}, I., {Taylor}, A.~M., {Semikoz}, D., \& {Neronov}, A. 2012, \apjl, 747,
  L14

\bibitem[{{Weidinger} \& {Spanier}(2010)}]{Weidinger:2010uq}
{Weidinger}, M., \& {Spanier}, F. 2010, \aap, 515, A18

\bibitem[{{Williams}(2005)}]{Williams:2005nx}
{Williams}, D.~A. 2005, in American Institute of Physics Conference Series,
  Vol. 745, High Energy Gamma-Ray Astronomy, ed. F.~A. {Aharonian}, H.~J.
  {V{\"o}lk}, \& D.~{Horns}, 499--504

\bibitem[{{Woo} {et~al.}(2005){Woo}, {Urry}, {van der Marel}, {Lira}, \&
  {Maza}}]{Woo:2005fk}
{Woo}, J.-H., {Urry}, C.~M., {van der Marel}, R.~P., {Lira}, P., \& {Maza}, J.
  2005, \apj, 631, 762

\end{thebibliography}

\end{document}